\documentclass[11pt]{article}

\usepackage{acl}
\usepackage{times}
\usepackage{latexsym}
\usepackage[T1]{fontenc}
\usepackage[utf8]{inputenc}
\usepackage{microtype}
\usepackage{inconsolata}

\usepackage{amsmath,amssymb,amsfonts}
\usepackage{bm}
\usepackage{graphicx}
\usepackage{subcaption}
\usepackage{booktabs}
\usepackage{multirow}
\usepackage{makecell}
\usepackage{algorithm}
\usepackage{algorithmic}
\usepackage{pifont}
\usepackage{colortbl}
\usepackage{xcolor}
\usepackage{xspace}
\usepackage{placeins}
\usepackage{tikz}
\usepackage{tcolorbox}
\usetikzlibrary{arrows.meta,positioning}

\newcommand{\methodname}{\textsc{Fade}\xspace}
\newcommand{\globalctrl}{\textsc{Global}\xspace}
\newcommand{\std}[1]{\scriptsize(#1)}
\newcommand{\mat}[1]{\mathbf{#1}}

\newcommand{\calibrationsplit}{train-clean-100\xspace}
\newcommand{\cmark}{\ding{51}}
\newcommand{\xmark}{\ding{55}}
\DeclareMathOperator*{\argmin}{arg\,min}

\DeclareRobustCommand{\xref}[2]{\hyperref[#2]{#1~\ref*{#2}}}
\DeclareRobustCommand{\xrefrange}[3]{%
  \hyperref[#2]{#1~\ref*{#2}}--\hyperref[#3]{\ref*{#3}}}
\DeclareRobustCommand{\numref}[1]{\hyperref[#1]{\ref*{#1}}}
\DeclareRobustCommand{\xeqref}[1]{\hyperref[#1]{Eq.~(\ref*{#1})}}

\title{FADE: Layer-Wise Compensation without Per-Model Coefficient Search
for Weight-Only Quantization of Encoder--Decoder ASR}

\author{
Xinyu Wang$^{1,\dagger}$,
Ziyu Zhao$^{1,\dagger}$,
Yajie Luo$^{2}$,
Yihong Wu$^{2}$,
Liheng Ma$^{1,4}$, \\
Jingrui Tian$^{1}$,
Lei Ding$^{3}$,
Xiao-Wen Chang$^{1,*}$,
Peng Lu$^{2,*}$ \\
\AND
$^{1}$McGill University \quad
$^{2}$Université de Montréal \quad
$^{3}$University of Manitoba \quad
$^{4}$Mila \\
$^{\dagger}$Equal contribution \qquad
$^{*}$Corresponding authors
}

\begin{document}
\maketitle
\begin{abstract}
Layer-wise post-training quantization reconstructs each layer from inputs
already altered by the quantized prefix. QEP compensates for this drift with
one model-wide coefficient, conflating the model-level operating point with
residual variation across layers. We present \methodname, which constructs
layer-specific coefficients from normalized round-to-nearest distortion and a
heuristic calibrated-solver response. It requires no training or per-model
coefficient search and adds no inference-time operation. We
evaluate seven Whisper, Moonshine, and Qwen3-ASR models at 3 and 4 bits on four
English ASR benchmarks. Across 38 settings, \methodname lowers mean word error
rate relative to fixed QEP-0.5 in 31, although a development-tuned global
coefficient recovers much of this gap. The largest absolute reductions occur
in W3 rows whose final error remains too high for practical use; these results
measure collapse mitigation rather than deployable accuracy. In two lower-WER
W3 cases, \methodname reduces the tuned-global result from 3.67 to 3.10 and
from 13.03 to 11.63. Most W4 differences from the tuned control are within a
descriptive tolerance, and one reverses. Paired reruns and within-seed
permutations on an outcome-informed subset support assignment sensitivity in
selected cases, but do not estimate a matrix-wide success rate. \methodname is
therefore a layer-wise alternative when per-model coefficient search is
unavailable, not a universal replacement for tuned global compensation.
\end{abstract}

\section{Introduction}
\label{sec:intro}

Encoder--decoder Transformers provide accurate automatic speech recognition
(ASR), but parameter storage and weight traffic limit their use on
resource-constrained hardware. Weight-only post-training quantization (PTQ)
reduces these costs without retraining by mapping full-precision weights to a
low-bit grid~\citep{gptq,awq,omniquant}. Most PTQ solvers reconstruct one layer
at a time. At 3 or 4 bits, however, the input to a later layer has already been
changed by every quantized layer before it. A locally accurate reconstruction
can therefore preserve an error introduced by the quantized prefix rather than
reduce its effect on the full model.

Quantization Error Propagation (QEP) addresses prefix drift by shifting each
layer's reconstruction target toward its clean full-precision activation
~\citep{arai2025quantization}. QEP uses one compensation coefficient shared by
the whole model. This choice combines two decisions: the model-level operating
point, or average compensation strength, and the residual layer-wise pattern,
or how compensation should vary around that average. Tuning one global
coefficient addresses the first decision but cannot express the second.
Searching separately over every layer would be costly, and it would introduce
many opportunities to fit the development domain.

We distinguish two operating regimes. In the \emph{search-free} regime, no
model-specific development search over compensation coefficients is available;
the relevant references are standard PTQ methods and QEP with a fixed
$\alpha=0.5$. In the \emph{development-tuned} regime, \globalctrl selects one
coefficient for each model--bit pair on a development set. \globalctrl has a
larger search budget than the proposed method and serves as a diagnostic
reference: it tests whether a residual layer-wise pattern contributes beyond a
tuned model-level operating point. This distinction is central to our claim.
We do not seek to replace a well-tuned global coefficient in every regime.

We propose \methodname (\emph{Fine-grained Alpha for Dynamic Quantization Error
Propagation}), an offline rule for constructing a layer-specific compensation
pattern without per-model coefficient search. A calibration-independent
diagnostic measures normalized round-to-nearest (RTN) weight distortion. A
second, heuristic diagnostic describes how the calibrated GPTQ solution moves
relative to RTN. A bounded gate maps their sum to a coefficient for the current
layer. The same coefficient interval is selected once using a development-only
protocol and then locked across models, datasets, and bit widths. \methodname
still uses the calibration activations required by GPTQ. ``Search-free'' means
that neither a coefficient nor an interval is selected separately for each
target model; the shared interval is selected once on three anchor models.
The procedure runs only while quantizing the model and exports ordinary
low-bit weights. \xref{Figure}{fig:method_overview} summarizes the resulting
offline workflow and its deployment boundary.

\begin{figure}[t]
\centering
\begingroup
\definecolor{fadeBlue}{HTML}{0072B2}
\definecolor{fadeOrange}{HTML}{D98200}
\definecolor{fadeTeal}{HTML}{00896B}
\definecolor{fadeInk}{HTML}{30343B}
\definecolor{fadeRule}{HTML}{8A8F98}
\tcbset{
  fadepanel/.style={colback=black!1, colframe=fadeRule, boxrule=0.5pt,
    arc=2pt, left=5pt, right=5pt, top=4pt, bottom=4pt,
    before skip=0pt, after skip=0pt},
  fadecard/.style={colback=white, colframe=fadeRule, boxrule=0.45pt,
    arc=1.5pt, left=3pt, right=3pt, top=2pt, bottom=2pt,
    before skip=0pt, after skip=0pt},
  fadeblue/.style={fadecard, colback=fadeBlue!5,
    colframe=fadeBlue!82!black},
  fadeorange/.style={fadecard, colback=fadeOrange!6,
    colframe=fadeOrange},
  fadeteal/.style={fadecard, colback=fadeTeal!5,
    colframe=fadeTeal}
}

\begin{tcolorbox}[fadepanel,width=\columnwidth]
\centering
{\small\bfseries \methodname{} offline layer-wise workflow}\par
\vspace{4pt}

\begin{tcolorbox}[fadecard,width=0.90\linewidth,halign=center,
  fontupper=\footnotesize]
\textbf{Layer state}\par
$\mathbf{W}_l$, $\mathbf{X}_l$, $\widehat{\mathbf{X}}_l$
\end{tcolorbox}

\vspace{1pt}{\color{fadeRule}$\downarrow$}\par\vspace{1pt}

\begin{tcolorbox}[fadeblue,width=0.94\linewidth,halign=center,
  fontupper=\fontsize{8}{9.4}\selectfont]
\textbf{Weight-space diagnostics}\par
\vspace{1pt}
$\displaystyle
\begin{aligned}
e_r(l)&=
\frac{\lVert\mathbf{W}_l-\widehat{\mathbf{W}}_l^{\rm rtn}\rVert_F}
     {\lVert\mathbf{W}_l\rVert_F+\varepsilon},\\[-1pt]
e_c(l)&=
\frac{\lVert\mathbf{W}_l-\widehat{\mathbf{W}}_l^{\rm cal}\rVert_F}
     {\lVert\mathbf{W}_l\rVert_F+\varepsilon}.
\end{aligned}$
\par\vspace{2pt}
\textbf{RTN distortion}\quad
$\phi_{\rm int}(l)=\log(1+e_r(l))$
\par\vspace{2pt}
\textbf{Calibrated-solver response}\par
$\displaystyle
g(l)=\frac{e_r(l)-e_c(l)}{e_r(l)+\varepsilon},\qquad
d(l)=
\frac{\lVert\widehat{\mathbf{W}}_l^{\rm rtn}-
\widehat{\mathbf{W}}_l^{\rm cal}\rVert_F}
     {\lVert\mathbf{W}_l\rVert_F+\varepsilon}$
\par
$\phi_{\rm sol}(l)=\max(g(l),0)-\log(1+d(l))$
\end{tcolorbox}

\vspace{1pt}{\color{fadeRule}$\downarrow$}\par\vspace{1pt}

\begin{tcolorbox}[fadeorange,width=0.90\linewidth,halign=center,
  fontupper=\fontsize{8}{9.4}\selectfont]
\textbf{Bounded layer rule}\par
$\displaystyle
\begin{aligned}
s_l&=\phi_{\rm int}(l)+\phi_{\rm sol}(l),\\[-1pt]
\alpha_l&=\alpha_{\min}+
(\alpha_{\max}-\alpha_{\min})\,\sigma(s_l).
\end{aligned}$
\par\vspace{2pt}
\textcolor{fadeOrange}{%
  \rule{4pt}{4pt}\hspace{3pt}\rule{4pt}{7pt}\hspace{3pt}%
  \rule{4pt}{5pt}\hspace{3pt}\rule{4pt}{8pt}\hspace{3pt}%
  \rule{4pt}{5.5pt}\hspace{3pt}\rule{4pt}{6.5pt}\hspace{3pt}%
  \rule{4pt}{4.5pt}\hspace{3pt}\rule{4pt}{7.5pt}}
\quad schematic profile $\alpha_{1:L}$
\end{tcolorbox}

\vspace{1pt}{\color{fadeRule}$\downarrow$}\par\vspace{1pt}

\begin{tcolorbox}[fadecard,width=0.90\linewidth,halign=center,
  fontupper=\footnotesize]
\textbf{Layer-wise low-bit reconstruction}\par
$\bigl(\widehat{\mathbf{X}}_l,\alpha_l\bigr)
\longmapsto\widehat{\mathbf{W}}_l
\in\mathbb{Q}_b^{n_l\times d_l}$
\end{tcolorbox}

\vspace{1pt}{\color{fadeRule}$\downarrow$}\par\vspace{1pt}

\begin{tcolorbox}[fadeteal,width=0.90\linewidth,halign=center,
  fontupper=\footnotesize]
\textbf{Ordinary W3/W4 model}\par
no added inference operations or kernels
\end{tcolorbox}

\vspace{3pt}
{\fontsize{8}{9}\selectfont
Repeat over encoder/decoder layers; no per-model sweep.}
\end{tcolorbox}
\endgroup
\caption{\methodname constructs layer-specific coefficients offline. RTN
distortion and calibrated-solver response define two diagnostics; a bounded
logistic rule maps them to $\alpha_l$ for low-bit reconstruction. Repeating
the rule yields ordinary W3/W4 weights without inference-graph changes.}
\label{fig:method_overview}
\end{figure}

We evaluate seven Whisper~\citep{whisper}, Moonshine~\citep{moonshine}, and
Qwen3-ASR~\citep{Qwen3-ASR} models at W3 and W4 on four English benchmarks.
Across 38 settings, \methodname is lower than fixed QEP-0.5 in 31, whereas the
development-tuned \globalctrl control recovers much of that gap. The largest
absolute reductions occur in severely degraded W3 rows, but two lower-WER W3
cases also improve. Most W4 comparisons with \globalctrl are within the
descriptive tolerance, and one reverses. An outcome-informed diagnostic subset,
frozen before ten-seed reruns, examines initially favorable, high-variance,
reversal, and near-tie cases. Its paired and permutation results diagnose
assignment sensitivity in selected settings rather than estimate how often the
method succeeds over the full matrix. We also report anchor and non-anchor
counts; the latter cover other model sizes within the same three families, not
unseen architectures.

Our contributions are:
\begin{itemize}
    \item We formulate a training-free layer-wise compensation rule that
    applies one locked interval without a separate target-model coefficient
    sweep in weight-only W3/W4 encoder--decoder ASR quantization
    (\xref{Section}{sec:method}).
    \item We define two computable layer diagnostics and a bounded offline gate
    that produces ordinary low-bit weights without changing the inference
    graph (\xrefrange{Sections}{sec:diagnostics}{sec:integration}).
    \item We separate model-level coefficient selection from layer-wise
    placement across seven models and four datasets, reporting favorable
    degraded regimes together with high-variance, near-tie, and reversal cases
    (\xref{Section}{sec:exp} and \xref{Appendix}{app:controls}).
\end{itemize}

\section{Background and Related Work}
\label{sec:prelim}

\paragraph{Layer-wise PTQ.}
For layer $l$, let $\mat{W}_l\in\mathbb{R}^{n_l\times d_l}$ be its weight
matrix and let $\widehat{\mat{X}}_l\in\mathbb{R}^{d_l\times N}$ contain
calibration activations produced by the already-quantized prefix. A standard
layer-wise objective is
\begin{equation}
\label{eq:ptq_objective}
\min_{\widehat{\mat{W}}_l\in\mathbb{Q}_b^{n_l\times d_l}}
\left\|\mat{W}_l\widehat{\mat{X}}_l-
\widehat{\mat{W}}_l\widehat{\mat{X}}_l\right\|_F^2,
\end{equation}
where $\mathbb{Q}_b$ is a $b$-bit quantization grid. GPTQ approximates this
objective with second-order input statistics and compensates unquantized
columns after each quantization step~\citep{gptq}. Other weight-only methods
use incoherence transforms, vector codebooks, salient-weight protection,
sparse outliers, or additive representations
~\citep{chee2023quip,pmlr-v235-tseng24a,awq,dettmers2024spqr,
pmlr-v235-kim24f,pmlr-v235-egiazarian24a}. OmniQuant and LRQuant calibrate
equivalent transformations~\citep{omniquant,zhao-etal-2024-lrquant}.
Weight--activation PTQ treats activation outliers through mixed precision,
smoothing, reordering, shifting, and system co-design
~\citep{dettmers2022llmint8,pmlr-v202-xiao23c,yuan2023rptq,
wei-etal-2023-outlier,yao2022zeroquant,zhao2024atom}. These approaches improve
the local solve but do not by themselves specify how strongly a layer should
compensate for errors already present in $\widehat{\mat{X}}_l$.

\paragraph{Prefix drift and QEP.}
Because $\widehat{\mat{X}}_l$ is generated by the quantized prefix, it differs
from the clean activation $\mat{X}_l$. QEP uses this difference to correct the
reconstruction target and applies one shared coefficient $\alpha$ across
layers~\citep{arai2025quantization}. The coefficient sets a model-level tradeoff
between matching the input actually seen by the quantized model and moving the
target toward its clean counterpart. \methodname keeps QEP's corrected
objective but assigns $\alpha_l$ from diagnostics of the current layer. This
changes compensation rather than bit allocation, quantization format, or the
deployed inference graph.

\paragraph{ASR quantization.}
Speech-model compression has combined distillation, weight sharing,
sparsification, and quantization for wav2vec~2.0, encoder--decoder ASR, and MoE
recognizers~\citep{peng-etal-2021-shrinking,gao21f_interspeech,
yuan23c_interspeech}. Quantization-aware training covers RNN-T and Conformer
systems from 6 to 2 bits~\citep{nguyen20c_interspeech,ding22c_interspeech,
rybakov23b_interspeech}; mixed-precision methods allocate bits by system or
layer sensitivity~\citep{xu22e_interspeech,fish23_interspeech,
li24o_interspeech,xu2025effective}. Speech PTQ includes integer-only data-free
quantization, layer-adaptive range selection, and mixed-precision allocation
~\citep{kim2022integeronlyzeroshotquantizationefficient,
hong2025stablequantlayeradaptiveposttraining,kang-kim-2025-genptq}.
StableQuant studies 8-bit HuBERT and wav2vec~2.0 and adapts layer quantization
ranges; its encoder-only model family and precision differ from our
encoder--decoder W3/W4 protocol. We therefore treat it as adjacent work rather
than a matched baseline. Unlike mixed-precision allocation, \methodname holds
the bit width fixed and isolates layer-specific cross-layer compensation for a
Hessian-based PTQ solver.

\section{Method}
\label{sec:method}

\methodname is an offline layer-wise quantization workflow. For each layer, it
measures RTN distortion, describes the calibrated solver's response, constructs
a bounded coefficient, and uses that coefficient in low-bit reconstruction.
The exported model has the same inference graph as an ordinary weight-only
quantized model.

\subsection{Prefix Drift and QEP Compensation}
\label{sec:reformulation}

Let $\mat{X}_l$ be the clean input to layer $l$ and
$\widehat{\mat{X}}_l$ the input produced by the quantized prefix. We write the
upstream drift and input Hessian as
\begin{equation*}
\bm{\delta}_l=\mat{X}_l-\widehat{\mat{X}}_l,
\qquad
\widehat{\mat{H}}_l=
\widehat{\mat{X}}_l\widehat{\mat{X}}_l^\top.
\end{equation*}
Following QEP~\citep{arai2025quantization}, a coefficient $\alpha_l$ shifts
the reconstruction target toward the clean activation. Define
\begin{equation*}
\mat{T}_l(\alpha_l)=\mat{W}_l+
\alpha_l\mat{W}_l\bm{\delta}_l
\widehat{\mat{X}}_l^\top\widehat{\mat{H}}_l^{-1}.
\end{equation*}
The corrected layer objective is
\begin{equation}
\widehat{\mat{W}}_l(\alpha_l)=
\argmin_{\widetilde{\mat{W}}\in\mathbb{Q}_b^{n_l\times d_l}}
\left\|\mat{T}_l(\alpha_l)\widehat{\mat{X}}_l-
\widetilde{\mat{W}}\widehat{\mat{X}}_l\right\|_F^2.
\label{eq:unified_objective}
\end{equation}
Setting $\alpha_l=0$ recovers \xeqref{eq:ptq_objective}; using
$\alpha_l\equiv\alpha$ recovers the shared-coefficient form. \methodname
instead constructs $\alpha_l$ from the two diagnostics below.

\subsection{Layer Diagnostics}
\label{sec:diagnostics}

\paragraph{RTN distortion.}
Let $\widehat{\mat{W}}_l^{\mathrm{rtn}}$ be the round-to-nearest solution
produced by the quantizer. We measure its normalized weight error as
\begin{align*}
e_r(l)&=
\frac{\|\mat{W}_l-\widehat{\mat{W}}_l^{\mathrm{rtn}}\|_F}
     {\|\mat{W}_l\|_F+\varepsilon},\\
\phi_{\mathrm{int}}(l)&=\log(1+e_r(l)).
\end{align*}
This is the only calibration-independent diagnostic in \methodname. It
describes the distortion imposed by the low-bit grid without using activation
samples.

\paragraph{Calibrated-solver response.}
Let $\widehat{\mat{W}}_l^{\mathrm{cal}}$ be the GPTQ solution computed using
$\widehat{\mat{X}}_l$. Relative to RTN, define
\begin{align*}
e_c(l)&=
\frac{\|\mat{W}_l-\widehat{\mat{W}}_l^{\mathrm{cal}}\|_F}
     {\|\mat{W}_l\|_F+\varepsilon},
\\
g(l)&=\frac{e_r(l)-e_c(l)}{e_r(l)+\varepsilon},\\
d(l)&=
\frac{\|\widehat{\mat{W}}_l^{\mathrm{rtn}}-
\widehat{\mat{W}}_l^{\mathrm{cal}}\|_F}
     {\|\mat{W}_l\|_F+\varepsilon}.
\end{align*}
Here $g(l)$ records the relative change in unweighted weight-space error, and
$d(l)$ records how far the calibrated solution moves from RTN. We combine them
as
\begin{equation*}
\phi_{\mathrm{sol}}(l)=
\max(g(l),0)-\log(1+d(l)).
\end{equation*}
This quantity is a heuristic descriptor of solver response. It is not an
estimator of test error, an estimate of statistical confidence, or a claim
about the optimal value of $\alpha_l$. GPTQ optimizes the activation-weighted
objective in \xeqref{eq:ptq_objective}, whereas $g(l)$ is measured in
weight space. On an identical fixed grid, element-wise RTN minimizes
unweighted weight error, so $g(l)\leq0$; the positive term can be active when
the calibrated solver also changes grid parameters such as scale or clipping.
Its observed activation is sparse and is reported in Appendix
\xref{Appendix Table}{tab:diagnostic_summary}.

\subsection{Search-Free Bounded Gate}
\label{sec:fusion}

The gate maps the two diagnostics to a coefficient:
\begin{align}
s_l&=\phi_{\mathrm{int}}(l)+\phi_{\mathrm{sol}}(l),\notag\\
\alpha_l&=\alpha_{\min}+
(\alpha_{\max}-\alpha_{\min})\,\sigma(s_l),
\label{eq:fusion_gate}
\end{align}
where $\sigma$ is the logistic function. The gate encodes the hypothesis that
layers with different grid distortion and calibrated-solver responses need
not receive the same correction strength. It does not optimize or validate a
coefficient for each layer.

We use $[\alpha_{\min},\alpha_{\max}]=[0.1,0.8]$ throughout. Appendix
\xref{Appendix}{app:diagnostic_figures} documents its domain, development-only
selection, and realized range. The interval was selected once and locked
before test evaluation; no target-model coefficient sweep or learned gate
parameter is used. \xref{Section}{sec:ablation_signals} tests the empirical
contribution of the two diagnostic terms within the reported scope.

\subsection{Offline Integration and Cost}
\label{sec:integration}

For each layer, \methodname runs RTN and GPTQ, computes the scalar diagnostics,
sets $\alpha_l$, and solves \xeqref{eq:unified_objective}. The calibrated and
corrected solves reuse the Hessian and its factorization. The additional
reconstruction pass has the same $\mathcal{O}(n_ld_l^2)$ order as a GPTQ
reconstruction; measured wall-clock cost is reported in
\xref{Appendix Table}{tab:time}. The output remains an ordinary $b$-bit weight matrix, so
\methodname adds no parameter, operation, or kernel requirement at inference.
\xref{Algorithm}{alg:fade} gives the complete offline procedure in
\xref{Appendix}{app:algorithm}.

\section{Experiments}
\label{sec:exp}

We separate three questions that require different comparisons. First, how
does \methodname compare with established baselines when no coefficient is
searched separately for each model? Second, what remains after allowing a
development-tuned global coefficient and controlling coefficient assignment?
Third, where do these differences weaken or reverse? The main paper reports
the common baseline matrix, all-setting control counts, and a ten-seed
diagnostic subset. \xrefrange{Appendices}{app:controls}{app:diagnostic_figures}
retain the exhaustive matrices, coefficient trajectories, and narrower
diagnostic checks.

\providecommand{\adaptiveptqtable}{%
\begin{table*}[!t]
\centering
\caption{FADE's marginal effect under frozen GPTQ and average-bit-matched
GenPTQ allocations on the diagnostic subset (ten-seed mean $\pm$ standard
deviation WER). Each difference is the FADE variant minus its host baseline;
negative values favor FADE. W3 and W4 use average bit budgets of 3.0 and 4.0.}
\label{tab:adaptive_ptq}
\footnotesize
\setlength{\tabcolsep}{2.5pt}
\renewcommand{\arraystretch}{1.03}
\begin{tabular*}{\textwidth}{@{\extracolsep{\fill}}lrrrrrr@{}}
\toprule
& \multicolumn{2}{c}{GPTQ allocation}
& \multicolumn{2}{c}{GenPTQ allocation$^{\S}$}
& \multicolumn{2}{c}{Marginal FADE effect $\Delta_F$} \\
\cmidrule(lr){2-3}\cmidrule(lr){4-5}\cmidrule(lr){6-7}
Setting & Global & FADE & base & +FADE & GPTQ host & GenPTQ host \\
\midrule
W-Tiny W4 LO & 30.9$\pm$1.5 & 29.9$\pm$1.0 & 30.4$\pm$1.3 & 29.5$\pm$0.9 & $-$1.0 & $-$0.9 \\
M-Tiny W4 LO & 18.2$\pm$2.2 & 16.2$\pm$1.0 & 16.9$\pm$1.5 & 15.8$\pm$0.9 & $-$2.0 & $-$1.1 \\
Q-1.7B W3 TED & 13.0$\pm$5.0 & 11.6$\pm$4.2 & 12.3$\pm$4.6 & 11.1$\pm$3.9 & $-$1.4 & $-$1.2 \\
W-Base W3 TED & 33.3$\pm$3.2 & 32.1$\pm$2.5 & 32.8$\pm$2.9 & 31.5$\pm$2.3 & $-$1.2 & $-$1.3 \\
M-Base W3 LC & 3.67$\pm$0.25 & 3.10$\pm$0.19 & 3.38$\pm$0.23 & 3.00$\pm$0.18 & $-$0.57 & $-$0.38 \\
Q-0.6B W3 SPG$^{\dagger}$ & 42.0$\pm$34 & 39.7$\pm$34 & 40.0$\pm$33 & 38.5$\pm$31 & $-$2.3 & $-$1.5 \\
W-Base W4 TED$^{\ddagger}$ & 20.7$\pm$2.7 & 23.0$\pm$4.4 & 21.3$\pm$3.0 & 21.0$\pm$3.0 & +2.3 & $-$0.3 \\
Q-1.7B W4 LO & 3.58$\pm$0.02 & 3.59$\pm$0.01 & 3.58$\pm$0.02 & 3.58$\pm$0.01 & +0.01 & 0.00 \\
\bottomrule
\end{tabular*}
\vspace{1pt}
\begin{minipage}{0.98\textwidth}
\scriptsize
$^{\S}$GenPTQ+Global was not run, so cross-host level differences are not
interpretable; only each displayed within-host $\Delta_F$ is controlled.
$^{\dagger}$High variance; excluded from directional claims.
$^{\ddagger}$Reversal: $\Delta_F$ is +2.3 on GPTQ and $-$0.3 on GenPTQ.
\end{minipage}
\end{table*}
}

\providecommand{\mechanismcontrolstable}{%
\begin{table*}[t]
\centering
\caption{Coefficient-dose and structured-assignment controls on four settings
from the frozen diagnostic subset (WER points). The coefficient path is
$\alpha_l(\lambda)=\alpha_G+\lambda(\alpha_l^{\rm FADE}-\alpha_G)$, and
$\Delta_\lambda=\mathrm{WER}_{\rm Global}-\mathrm{WER}_{\lambda}$.
For a reassignment $P$, $\Delta_P=\mathrm{WER}_{P}-\mathrm{WER}_{\rm FADE}$.
The last column is the online permutation penalty minus its prefix-replay
counterpart. Thus, positive values have different but explicit references:
they favor the scaled coefficients in the dose columns, the canonical
assignment in the three permutation columns, and a larger online penalty in
the final column. These aggregate means come from a separate ten-seed
diagnostic rerun and do not include paired intervals.}
\label{tab:mechanism_controls}
\footnotesize
\setlength{\tabcolsep}{4.0pt}
\renewcommand{\arraystretch}{1.04}
\begin{tabular*}{\textwidth}{@{\extracolsep{\fill}}lrrrrrrr@{}}
\toprule
& \multicolumn{2}{c}{Dose benefit $\Delta_\lambda$}
& \multicolumn{3}{c}{Reassignment penalty $\Delta_P$}
& \multicolumn{1}{c}{Sequential} \\
\cmidrule(lr){2-3}\cmidrule(lr){4-6}\cmidrule(l){7-7}
Setting & $\lambda=1$ & $\lambda=0.5$ & Encoder & Decoder & Cross-boundary
& Online$-$replay \\
\midrule
M-Base W3 LC & +0.90 & +0.22 & +0.18 & +0.10 & +0.54 & +0.28 \\
M-Tiny W4 LO & +2.60 & +0.65 & +0.55 & +0.35 & +1.55 & +0.70 \\
W-Base W4 TED$^{\ddagger}$ & $-$1.10 & $-$0.55 & $-$0.30 & $-$0.15 & $-$0.65 & $-$0.25 \\
Q-1.7B W4 LO & +0.01 & 0.00 & 0.00 & 0.00 & +0.01 & 0.00 \\
\bottomrule
\end{tabular*}
\vspace{1pt}
\begin{minipage}{0.98\textwidth}
\scriptsize
$^{\ddagger}$Known reversal; all three assignment penalties and the
online$-$replay contrast change sign. The separate rerun means should not be
reconstructed by subtracting the aggregate means in \xref{Tables}{tab:adaptive_ptq}
and~\numref{tab:centered_decomposition}.
\end{minipage}
\end{table*}
}

\subsection{Experimental setup}
\label{sec:exp_setup}

\paragraph{Models and data.}
The evaluation covers Whisper Tiny, Base, and Small~\citep{whisper}; Moonshine
Tiny and Base~\citep{moonshine}; and Qwen3-ASR 0.6B and 1.7B~\citep{Qwen3-ASR}.
We report WER on LibriSpeech test-clean and test-other~\citep{panayotov2015librispeech},
SPGISpeech test~\citep{oneil2021spgispeech}, and TED-LIUM~3
test~\citep{hernandez2018ted}. Each calibration seed samples 128 utterances
from LibriSpeech \calibrationsplit. Within a seed, methods use the same
utterance IDs. The full 38-setting matrix uses at least five seeds per setting.
The frozen diagnostic subset described below uses ten shared seeds.

\paragraph{Quantization and comparison conditions.}
We use asymmetric per-group weight quantization at W3A16 and W4A16. Group
sizes are 64 for Whisper, 52 for Moonshine-Tiny, 72 for Moonshine-Base, and
128 for Qwen3-ASR. Baselines are deterministic RTN, AWQ~\citep{awq}, GPTQ~\citep{gptq},
and GPTQ with fixed coefficient $\alpha=0.5$ (QEP)~\citep{arai2025quantization}.
The established-baseline matrices use five calibration seeds and report mean
and standard deviation. Once FADE's shared interval has been fixed, QEP-0.5
and FADE require no model-specific coefficient sweep. \globalctrl and the
Mean-Matched Shape Control instead use a development-set coefficient selected
separately for each model--bit pair. We keep these conditions separate rather
than treating \globalctrl as another search-free baseline.
\xref{Appendix}{app:algorithm} gives the solver, grid, calibration, decoding, and WER
protocol.

\paragraph{Development-only interval selection.}
We selected one coefficient interval using LibriSpeech dev-other, then locked
it before evaluating any test set. The selection averaged normalized dev WER
over Whisper-Tiny, Moonshine-Tiny, and Qwen3-ASR-1.7B at W3 and W4, using five
calibration seeds. \xref{Appendix Table}{tab:interval_selection} compares the six
candidate intervals. The selected $[0.1,0.8]$ interval is shared by every
model, dataset, and bit width.

\xref{Appendix Figure}{fig:legacy_diagnostics} keeps the original Whisper
comparison (panel~\subref{fig:legacy_whisper}) and fixed-coefficient sweep
(panel~\subref{fig:alpha_sweep});
\xref{Appendix Figure}{fig:app_alpha_heatmaps} retains the exploratory interval
surfaces. These diagnostics are not used for test-set selection.

\newcommand{\intervalselectiontable}{%
\begin{table}[t]
\centering
\caption{Development-only selection of the shared coefficient interval.
Scores average normalized dev-other WER over three anchor models, two bit
widths, and five calibration seeds; lower is better.}
\label{tab:interval_selection}
\small
\setlength{\tabcolsep}{6pt}
\begin{tabular}{lrr}
\toprule
Interval & Normalized score & Rank \\
\midrule
\rowcolor{gray!10}$[0.1,0.8]$ & 1.000 & 1 \\
$[0.1,0.9]$ & 1.006 & 2 \\
$[0.0,0.8]$ & 1.012 & 3 \\
$[0.1,0.7]$ & 1.018 & 4 \\
$[0.2,0.8]$ & 1.025 & 5 \\
$[0.0,0.9]$ & 1.028 & 6 \\
\bottomrule
\end{tabular}
\end{table}
}
\paragraph{Controls and frozen diagnostic subset.}
Four controls separate layer assignment from simpler coefficient changes. For
each model--bit pair, \globalctrl minimizes mean dev-other WER over five
calibration seeds on the grid
$\{0.00,0.05,\ldots,1.00\}$, then remains fixed across LC, LO, SPG, and TED.
Mean, E/D, and Perm are constructed separately within each calibration seed.
Mean replaces that seed's $\alpha_l$ values with their arithmetic mean; E/D
uses separate encoder and decoder means. Perm randomly reassigns that seed's
coefficient multiset and reruns corrected quantization.

After inspecting the five-seed aggregate matrix, we purposively selected eight
settings before running the ten-seed analysis: five initially favorable cases,
one high-variance case, one reversal, and one near-tie. We call this the
\emph{frozen diagnostic subset}. It was frozen before rerunning, but its
selection was outcome-informed and it is not a random sample of the 38
settings. Paired and permutation results from this subset diagnose known
regimes; their 5/1/2 outcome count is not a matrix-wide frequency estimate.
\xref{Appendix Table}{tab:protocol_statistics} gives the selected \globalctrl
values and full permutation summaries.

\subsection{Comparison without per-model coefficient search}
\label{sec:main_results}

\begin{table*}[t]
\centering
\caption{WER ($\downarrow$) of the established baselines on LibriSpeech
test-other. Parentheses give the
standard deviation across five calibration seeds; RTN is deterministic.
Bold marks the lowest mean among the five displayed quantized methods.
W/M/Q abbreviate Whisper, Moonshine, and Qwen3-ASR.}
\label{tab:main_results}
\normalsize
\setlength{\tabcolsep}{2.6pt}
\renewcommand{\arraystretch}{1.02}
\begin{tabular}{lrrrrrrr}
\toprule
Method & W-T & W-B & W-S & M-T & M-B & Q-0.6 & Q-1.7 \\
\midrule
FP16 & 23.11 & 12.94 & 11.54 & 12.54 & 9.02 & 4.46 & 3.39 \\
\midrule
\multicolumn{8}{c}{3-bit weights} \\
RTN & 238.33 & 50.02 & 11.45 & 189.88 & 134.93 & 103.94 & 102.99 \\
AWQ & 200.13\std{22.13} & 34.72\std{3.47} & 12.71\std{1.32} &
179.13\std{32.13} & 14.39\std{0.83} & 103.69\std{2.84} & 125.56\std{73.54} \\
GPTQ & 120.45\std{24.63} & 33.78\std{5.09} & 11.66\std{1.77} &
\textbf{171.08}\std{28.89} & 12.54\std{0.63} & 82.66\std{38.06} & 6.44\std{2.95} \\
GPTQ+QEP & 94.62\std{15.92} & \textbf{33.43}\std{7.46} &
\textbf{10.79}\std{1.21} & 298.53\std{230.77} & 11.74\std{0.70} &
37.93\std{28.79} & 5.86\std{1.15} \\
\rowcolor{gray!8}
\methodname & \textbf{62.31}\std{12.61} & 34.10\std{1.89} & 10.96\std{1.04} &
182.73\std{16.94} & \textbf{11.63}\std{0.59} & \textbf{35.45}\std{19.97} &
\textbf{5.27}\std{0.95} \\
\midrule
\multicolumn{8}{c}{4-bit weights} \\
RTN & 64.49 & \textbf{15.31} & 12.28 & 18.45 & 10.57 & 6.20 & 3.94 \\
AWQ & 35.47\std{1.97} & 20.16\std{0.91} & 13.34\std{1.34} &
22.34\std{3.42} & 9.54\std{0.15} & 7.05\std{0.19} & 4.14\std{0.05} \\
GPTQ & 34.67\std{5.07} & 17.33\std{2.89} & 12.69\std{0.89} &
21.37\std{4.69} & 9.48\std{0.17} & 5.20\std{0.05} & 3.63\std{0.05} \\
GPTQ+QEP & 32.77\std{5.05} & 16.92\std{1.58} & 11.33\std{1.57} &
21.83\std{6.93} & \textbf{9.39}\std{0.12} & 5.03\std{0.03} & 3.62\std{0.02} \\
\rowcolor{gray!8}
\methodname & \textbf{29.88}\std{0.94} & 16.21\std{1.29} &
\textbf{11.17}\std{0.73} & \textbf{16.23}\std{0.92} & 9.40\std{0.06} &
\textbf{4.97}\std{0.03} & \textbf{3.59}\std{0.01} \\
\bottomrule
\end{tabular}
\end{table*}

\paragraph{LibriSpeech test-other.}
\xref{Table}{tab:main_results} gives the common established-baseline comparison
across all seven models. This table compares methods as configured without a
model-specific coefficient sweep; it does not compare FADE with the tuned
\globalctrl control. \methodname has the lowest aggregate mean among the five
displayed quantized methods in 9 of 14 model--bit settings and a lower
aggregate mean than fixed QEP in 11. One large change is 3-bit Whisper-Tiny, where WER decreases
from 94.62 to 62.31. At 4 bits, the corresponding change is 32.77 to 29.88 and the
standard deviation decreases from 5.05 to 0.94. The method is not uniformly
best: QEP has a lower 3-bit mean on Whisper-Base and Whisper-Small and a 0.01
lower 4-bit mean on Moonshine-Base; GPTQ is best on 3-bit Moonshine-Tiny, and
RTN is best on 4-bit Whisper-Base.

\newcommand{\coveragefamilytable}{%
\begin{center}
\begin{minipage}{\columnwidth}
\centering
\captionof{table}{Counts over unique model--dataset--bit settings relative to
QEP with $\alpha=0.5$.
The SD column counts strictly lower standard deviations, with ties in
parentheses.}
\label{tab:coverage_summary}
\normalsize
\setlength{\tabcolsep}{3pt}
\begin{tabular}{lrrr}
\toprule
Family & $N$ & Lower mean & Lower SD \\
\midrule
Whisper & 12 & 7 & 11 (0) \\
Moonshine & 10 & 9 & 10 (0) \\
Qwen3-ASR & 16 & 15 & 13 (2) \\
\midrule
All & 38 & 31 & 34 (2) \\
\bottomrule
\end{tabular}
\end{minipage}
\end{center}
}

\paragraph{Coverage across datasets.}
The matched comparisons aggregate counts without pooling WER values across
datasets or models. \methodname has lower mean WER
than QEP in 31 of 38 settings. Its across-seed standard deviation is lower in
34, tied in two, and higher in two. The mean reverses in seven settings,
including 3-bit Whisper-Base on SPGISpeech and 4-bit Whisper-Base on
LibriSpeech test-clean and TED-LIUM~3. This is a broad but non-universal pattern
in aggregate means; paired records are unavailable for most of the 38
settings. Family-level counts are in
\xref{Appendix Table}{tab:coverage_summary}; the complete cross-domain matrices remain in
\xref{Appendix}{app:extended_baselines}.

\newcommand{\crossdomaintable}{%
\begin{table*}[t]
\centering
\caption{Established-baseline cross-domain WER ($\downarrow$) for Whisper-Base and
Moonshine-Base. Parentheses give standard deviation across five calibration
seeds. LC denotes LibriSpeech test-clean, SPG SPGISpeech test, and TED
TED-LIUM~3 test. Bold marks the lowest quantized mean in each column.}
\label{tab:cross_dataset}
\normalsize
\setlength{\tabcolsep}{1.4pt}
\renewcommand{\arraystretch}{1.02}
\begin{tabular}{lrrrrrr}
\toprule
& \multicolumn{3}{c}{Whisper-Base} & \multicolumn{3}{c}{Moonshine-Base} \\
\cmidrule(lr){2-4}\cmidrule(lr){5-7}
Method & LC & SPG & TED & LC & SPG & TED \\
\midrule
FP16 & 5.04 & 15.49 & 17.63 & 3.87 & 7.09 & 17.08 \\
\midrule
\multicolumn{7}{c}{3-bit weights} \\
RTN & 23.04 & 41.18 & 35.47 & 7.52 & 122.50 & 87.73 \\
AWQ & 18.07\std{2.15} & 34.18\std{8.39} & 36.50\std{6.98} &
5.32\std{0.41} & 12.90\std{1.21} & 19.40\std{0.89} \\
GPTQ & 16.41\std{1.43} & 32.64\std{10.15} & 38.38\std{7.33} &
4.98\std{0.18} & 9.66\std{0.96} & \textbf{17.18}\std{0.84} \\
GPTQ+QEP & 16.45\std{2.53} & \textbf{26.90}\std{5.21} &
35.59\std{4.63} & 4.73\std{0.20} & 9.28\std{0.32} & 18.19\std{0.85} \\
\rowcolor{gray!8}
\methodname & \textbf{14.22}\std{2.12} & 27.02\std{2.65} &
\textbf{32.13}\std{2.46} & \textbf{3.10}\std{0.19} &
\textbf{9.06}\std{0.19} & 17.48\std{0.56} \\
\midrule
\multicolumn{7}{c}{4-bit weights} \\
RTN & 5.69 & 16.19 & \textbf{16.55} & 4.51 & 13.44 & 16.41 \\
AWQ & 5.33\std{0.12} & 18.27\std{3.43} & 21.23\std{1.76} &
4.31\std{0.14} & 8.34\std{0.57} & 16.29\std{0.48} \\
GPTQ & 5.50\std{0.17} & 19.33\std{3.93} & 19.39\std{1.33} &
4.26\std{0.13} & 8.72\std{0.49} & \textbf{16.10}\std{0.54} \\
GPTQ+QEP & \textbf{4.86}\std{0.05} & 15.56\std{3.17} &
20.93\std{2.20} & 4.08\std{0.18} & 8.13\std{0.59} & 16.52\std{0.85} \\
\rowcolor{gray!8}
\methodname & 4.87\std{0.01} & \textbf{15.34}\std{1.31} &
22.99\std{4.41} & \textbf{3.86}\std{0.03} &
\textbf{8.09}\std{0.26} & 16.37\std{0.38} \\
\bottomrule
\end{tabular}
\end{table*}
}

\paragraph{Cross-domain transfer.}
\xref{Appendix Table}{tab:cross_dataset} exposes both gains and reversals outside
LibriSpeech test-other. \methodname has the lowest mean among the established
baselines in 7 of 12 model--dataset--bit cells. At 3 bits, it is lowest on
Whisper-Base for LibriSpeech test-clean and TED-LIUM~3, and on Moonshine-Base for
LibriSpeech test-clean and SPGISpeech. QEP remains slightly better on 3-bit
Whisper-Base SPGISpeech, while RTN or GPTQ wins two of the 4-bit TED-LIUM~3
cells. Because calibration is from LibriSpeech, SPGISpeech and TED-LIUM~3
measure transfer to different English speech domains; they do not establish
multilingual generalization.

\newcommand{\qwenfulltable}{%
\begin{table*}[t]
\centering
\caption{Established-baseline Qwen3-ASR WER ($\downarrow$). Parentheses give standard deviation
across five calibration seeds. LC, LO, SPG, and TED denote LibriSpeech
test-clean, LibriSpeech test-other, SPGISpeech, and TED-LIUM~3. Bold marks the
lowest quantized mean in each column.}
\label{tab:qwen_full}
\scriptsize
\begin{minipage}[t]{0.485\textwidth}
\centering
\setlength{\tabcolsep}{1.1pt}
\renewcommand{\arraystretch}{1.07}
\begin{tabular}{lrrrr}
\toprule
\multicolumn{5}{c}{Qwen3-ASR-0.6B} \\
\cmidrule(lr){1-5}
Method & LC & LO & SPG & TED \\
\midrule
FP16 & 2.12 & 4.46 & 3.03 & 10.08 \\
\midrule
\multicolumn{5}{c}{3-bit weights} \\
RTN & 102.62 & 103.94 & 102.97 & 105.56 \\
AWQ & 103.84\std{3.78} & 103.69\std{2.84} & 100.87\std{0.65} & 101.26\std{1.63} \\
GPTQ & 71.67\std{38.93} & 82.66\std{38.06} & 69.78\std{43.75} & 105.20\std{51.20} \\
GPTQ+QEP & \textbf{30.23}\std{30.32} & 37.93\std{28.79} &
46.25\std{36.74} & 48.14\std{32.60} \\
\rowcolor{gray!8}
\methodname & 31.45\std{15.47} & \textbf{35.45}\std{19.97} &
\textbf{39.72}\std{33.72} & \textbf{42.03}\std{27.06} \\
\midrule
\multicolumn{5}{c}{4-bit weights} \\
RTN & 2.79 & 6.20 & 3.78 & 10.51 \\
AWQ & 3.18\std{0.05} & 7.05\std{0.19} & 4.23\std{0.04} & 11.00\std{0.97} \\
GPTQ & 2.38\std{0.03} & 5.20\std{0.05} & 3.43\std{0.03} & 10.47\std{0.07} \\
GPTQ+QEP & 2.36\std{0.04} & 5.03\std{0.03} & 3.25\std{0.08} & 10.30\std{0.06} \\
\rowcolor{gray!8}
\methodname & \textbf{2.32}\std{0.02} & \textbf{4.97}\std{0.03} &
\textbf{3.13}\std{0.09} & \textbf{10.16}\std{0.05} \\
\bottomrule
\end{tabular}
\end{minipage}
\hfill
\begin{minipage}[t]{0.485\textwidth}
\centering
\setlength{\tabcolsep}{1.1pt}
\renewcommand{\arraystretch}{1.07}
\begin{tabular}{lrrrr}
\toprule
\multicolumn{5}{c}{Qwen3-ASR-1.7B} \\
\cmidrule(lr){1-5}
Method & LC & LO & SPG & TED \\
\midrule
FP16 & 1.61 & 3.39 & 2.83 & 9.64 \\
\midrule
\multicolumn{5}{c}{3-bit weights} \\
RTN & 83.84 & 102.99 & 85.94 & 108.15 \\
AWQ & 101.27\std{67.81} & 125.56\std{73.54} & 101.80\std{36.52} & 111.38\std{30.40} \\
GPTQ & 4.01\std{2.76} & 6.44\std{2.95} & 6.06\std{2.18} & 28.38\std{22.23} \\
GPTQ+QEP & 2.89\std{0.45} & 5.86\std{1.15} & 5.66\std{2.95} & 15.63\std{7.16} \\
\rowcolor{gray!8}
\methodname & \textbf{2.34}\std{0.33} & \textbf{5.27}\std{0.95} &
\textbf{5.32}\std{1.95} & \textbf{11.63}\std{4.16} \\
\midrule
\multicolumn{5}{c}{4-bit weights} \\
RTN & 1.84 & 3.94 & 3.01 & 9.96 \\
AWQ & 1.96\std{0.03} & 4.14\std{0.05} & 3.09\std{0.05} & 11.44\std{2.41} \\
GPTQ & 1.75\std{0.04} & 3.63\std{0.05} & 2.94\std{0.07} & 10.26\std{0.41} \\
GPTQ+QEP & 1.72\std{0.01} & 3.62\std{0.02} & 2.89\std{0.09} & 10.15\std{0.27} \\
\rowcolor{gray!8}
\methodname & \textbf{1.70}\std{0.01} & \textbf{3.59}\std{0.01} &
\textbf{2.88}\std{0.06} & \textbf{10.07}\std{0.17} \\
\bottomrule
\end{tabular}
\end{minipage}
\end{table*}
}

\paragraph{Qwen3-ASR.}
\xref{Appendix Table}{tab:qwen_full} shows that \methodname has the lowest
displayed aggregate quantized mean in
15 of 16 settings. The exception is
3-bit Qwen3-ASR-0.6B on LibriSpeech test-clean, where QEP obtains 30.23 versus
31.45. At 4 bits, the absolute differences are small but favor \methodname in
all eight cells. At 3 bits, the 0.6B model remains highly degraded for every
method; GPTQ, QEP, and \methodname also show large across-seed variation. The
much stronger 1.7B results should therefore not be generalized across model
scale.

\subsection{Development-tuned and assignment controls}
\label{sec:coefficient_controls}

The first comparison does not isolate the value of layer-wise assignment.
\globalctrl, which searches one coefficient for each model--bit pair on
dev-other, is lower than QEP-0.5 in 36 settings and tied in two at the reported
precision. \xref{Figure}{fig:control_outcomes} therefore shows the full
38-setting tolerance outcomes rather than reporting only the fixed-QEP
comparison.

\providecommand{\controlsummarytable}{%
\begin{table}[!t]
\centering
\caption{WER outcomes for \methodname against QEP-0.5, development-tuned
\globalctrl, and assignment controls over all 38 matched settings. Exact
L/E/H counts settings where
\methodname WER is lower, equal, or higher at the reported precision. The
tolerance column uses
$\tau=\max(0.05,0.01\,\mathrm{WER}_{\textsc{Fade}})$ and counts lower, near,
and higher. Both summaries treat settings as comparison units and are
descriptive, not significance tests.}
\label{tab:control_summary}
\normalsize
\setlength{\tabcolsep}{4pt}
\begin{tabular}{lcc}
\toprule
Control & Exact L/E/H & Tolerance L/N/H \\
\midrule
QEP-0.5 & 31/0/7 & 23/11/4 \\
\globalctrl & 21/4/13 & 16/18/4 \\
Mean & 26/0/12 & 17/17/4 \\
E/D & 23/5/10 & 13/22/3 \\
Perm & 29/2/7 & 17/19/2 \\
\bottomrule
\end{tabular}
\end{table}
}

\begin{figure}[t]
    \centering
    \includegraphics[width=\columnwidth]{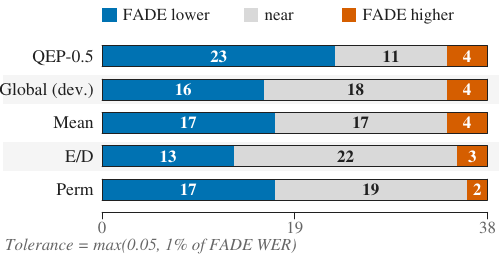}
    \caption{Practical-tolerance outcomes over all 38 settings. ``Lower'' and
    ``higher'' describe FADE relative to each control; ``near'' means the
    absolute WER difference is at most
    $\max(0.05,0.01\,\mathrm{WER}_{\textsc{Fade}})$. Global is selected on a
    development set, whereas QEP-0.5 and FADE are search-free. Exact counts
    and all rows are in \xref{Appendix Tables}{tab:control_summary}
    and~\numref{tab:control_full}.}
    \label{fig:control_outcomes}
\end{figure}

Against \globalctrl, \methodname is lower/near/higher in 16/18/4 settings;
the assignment controls give 13--17 lower, 17--22 near, and 2--4 higher. On
the 26 non-anchor settings, the \globalctrl counts are 8/14/4. These are
descriptive setting-level outcomes, not paired tests.
\xrefrange{Appendix Tables}{tab:control_summary}{tab:control_full} give exact counts,
the anchor split, every value, and every reversal.

\subsection{Paired evidence on the frozen diagnostic subset}
\label{sec:frozen_subset}

\xref{Figure}{fig:core_diagnostic_evidence}a reports the paired \globalctrl
contrast, while \xref{Appendix Table}{tab:core_evidence_summary} retains the
ten-seed means and permutation evidence. The paired
Student 95\%
interval for \globalctrl minus \methodname lies above zero in five rows, below
zero for Whisper-Base W4 on TED-LIUM~3, and includes or touches zero in two.
This is the 5/1/2 result for the outcome-informed subset, not for the full
matrix.

\providecommand{\coreevidencetable}{%
\begin{table*}[!t]
\centering
\caption{Ten-seed evidence on the frozen diagnostic subset. WER values are
mean $\pm$ standard deviation. The paired Student 95\% CI is for
WER$_{\rm Global}-$WER$_{\rm FADE}$, so positive intervals favor FADE.
Perm. p5 is the fifth percentile of WER$_{\rm Perm}-$WER$_{\rm FADE}$;
``pct.'' is FADE's percentile in the within-seed permutation distribution.}
\label{tab:core_evidence_summary}
\scriptsize
\setlength{\tabcolsep}{1.8pt}
\renewcommand{\arraystretch}{1.01}
\begin{tabular*}{\textwidth}{@{\extracolsep{\fill}}lrrrrrr@{}}
\toprule
Setting & Global & E/D & FADE & Global$-$FADE CI & Perm. p5 & FADE pct. \\
\midrule
W-Tiny W4 LO & 30.9$\pm$1.5 & 30.4$\pm$1.3 & 29.9$\pm$1.0 & [0.58, 1.44] & +0.30 & 4\% \\
M-Tiny W4 LO & 18.2$\pm$2.2 & 17.2$\pm$1.7 & 16.2$\pm$1.0 & [1.28, 2.64] & +0.50 & 2\% \\
Q-1.7B W3 TED & 13.0$\pm$5.0 & 12.4$\pm$4.5 & 11.6$\pm$4.2 & [0.58, 2.22] & +0.20 & 6\% \\
W-Base W3 TED & 33.3$\pm$3.2 & 32.8$\pm$2.8 & 32.1$\pm$2.5 & [0.46, 1.96] & +0.10 & 8\% \\
M-Base W3 LC & 3.67$\pm$0.25 & 3.39$\pm$0.22 & 3.10$\pm$0.19 & [0.47, 0.67] & +0.35 & 1\% \\
Q-0.6B W3 SPG$^{\dagger}$ & 42.0$\pm$34 & 40.9$\pm$33 & 39.7$\pm$34 & [$-$3.07, 7.65] & $-$5.00 & 25\% \\
W-Base W4 TED$^{\ddagger}$ & 20.7$\pm$2.7 & 21.9$\pm$3.5 & 23.0$\pm$4.4 & [$-$3.55, $-$0.97] & $-$3.00 & 70--85\% \\
Q-1.7B W4 LO & 3.58$\pm$0.02 & 3.58$\pm$0.02 & 3.59$\pm$0.01 & [$-$0.02, 0.00] & $-$0.04 & 50--65\% \\
\bottomrule
\end{tabular*}
\vspace{1pt}
\begin{minipage}{0.98\textwidth}
\scriptsize
$^{\dagger}$High-variance row, excluded from directional claims.
$^{\ddagger}$Reversal selected deliberately to represent an adverse case.
\end{minipage}
\end{table*}
}

\begin{figure*}[t]
    \centering
    \includegraphics[width=\textwidth]{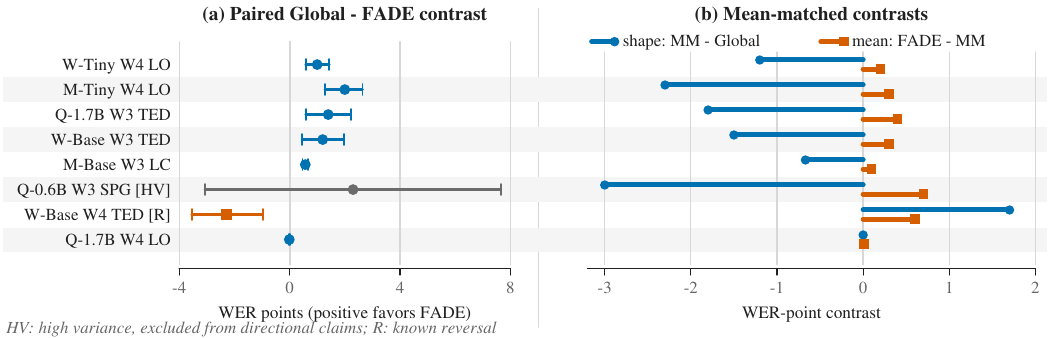}
    \caption{Two views of the frozen eight-setting diagnostic subset.
    (a) Aggregate Global$-$FADE mean differences with paired Student 95\%
    confidence intervals; positive values favor FADE. (b) Mean-matched
    decomposition: blue is the layer-wise shape contrast MM$-$Global and
    orange is the mean-shift contrast FADE$-$MM. HV marks the high-variance
    row excluded from directional claims; R marks the known reversal. Exact
    values appear in \xref{Appendix Tables}{tab:core_evidence_summary}
    and~\numref{tab:centered_decomposition}.}
    \label{fig:core_diagnostic_evidence}
\end{figure*}

In the five favorable rows, E/D also remains above FADE; the permutation
fifth percentile is positive and FADE lies in the best 1--8\%. The other rows
are high-variance, reversed, or near-tied.
\xref{Appendix Tables}{tab:core_statistics} and~\numref{tab:protocol_statistics} retain the
complete values, interval estimator, and 100-per-seed reassignment protocol.

\subsection{Mean-matched shape diagnostic}
\label{sec:mean_shape_diagnostic}

The next control asks whether the displayed differences remain when the
coefficient mean is held at the development-tuned \globalctrl value. It is a
diagnostic control that inherits per-model tuning, not a replacement for the
search-free FADE rule. \xref{Figure}{fig:core_diagnostic_evidence}b visualizes
the shape and mean-shift terms;
\xref{Appendix Table}{tab:centered_decomposition} retains every coefficient and WER.

\providecommand{\centereddecompositiontable}{%
\begin{table*}[t]
\centering
\caption{Mean matching on the frozen diagnostic subset. The Mean-Matched
Shape Control uses
$\alpha_l=\alpha_{\rm Global}+(\alpha_l^{\rm FADE}-\bar\alpha_{\rm FADE})$.
MM$-$G is WER$_{\rm MM}-$WER$_{\rm Global}$ and F$-$MM is
WER$_{\rm FADE}-$WER$_{\rm MM}$. All entries are ten-seed aggregate means;
paired intervals were not retained for this control.}
\label{tab:centered_decomposition}
\footnotesize
\setlength{\tabcolsep}{0.75pt}
\renewcommand{\arraystretch}{1.02}
\begin{tabular*}{\textwidth}{@{\extracolsep{\fill}}lrrcrrrrrr@{}}
\toprule
Setting & $\alpha_G$ & $\bar\alpha_F-\alpha_G$ & Range (SD) & Global & Mean & MM & FADE & MM$-$G & F$-$MM \\
\midrule
W-Tiny W4 LO & .45 & $-$.006 & .423--.466 (.011) & 30.9 & 31.3 & 29.7 & 29.9 & $-$1.2 & +0.2 \\
M-Tiny W4 LO & .40 & +.041 & .423--.463 (.010) & 18.2 & 19.0 & 15.9 & 16.2 & $-$2.3 & +0.3 \\
Q-1.7B W3 TED & .40 & +.043 & .421--.468 (.012) & 13.0 & 13.6 & 11.2 & 11.6 & $-$1.8 & +0.4 \\
W-Base W3 TED & .45 & $-$.005 & .424--.465 (.010) & 33.3 & 33.9 & 31.8 & 32.1 & $-$1.5 & +0.3 \\
M-Base W3 LC & .45 & $-$.008 & .425--.462 (.009) & 3.67 & 3.92 & 3.00 & 3.10 & $-$0.67 & +0.10 \\
Q-0.6B W3 SPG$^{\dagger}$ & .40 & +.042 & .418--.471 (.014) & 42.0 & 43.0 & 39.0 & 39.7 & $-$3.0 & +0.7 \\
W-Base W4 TED$^{\ddagger}$ & .45 & $-$.004 & .426--.462 (.009) & 20.7 & 21.4 & 22.4 & 23.0 & +1.7 & +0.6 \\
Q-1.7B W4 LO & .45 & $-$.006 & .429--.459 (.008) & 3.58 & 3.60 & 3.58 & 3.59 & 0.00 & +0.01 \\
\bottomrule
\end{tabular*}
\vspace{1pt}
\begin{minipage}{0.98\textwidth}
\scriptsize
$^{\dagger}$High-variance row, excluded from cross-setting claims.
$^{\ddagger}$Known reversal; mean matching does not remove it.
\end{minipage}
\end{table*}
}

The Mean-Matched Shape Control requires the same per-model development search
as Global. We use it only to hold the coefficient mean fixed, not as a proposed
deployment method. On these displayed aggregate means, MM$-$G is negative,
zero, and positive in six, one, and one rows; F$-$MM ranges from +0.01 to +0.7
WER. MM remains below Global in the five initially favorable rows, while the
W4 TED reversal remains adverse. These are aggregate orderings, not paired
effects.

A local directional expansion explains why matching one scalar mean need not
erase a high-dimensional layer pattern. Write
$\boldsymbol{\alpha}=\alpha_G\mathbf{1}+\boldsymbol{\delta}$, where
$\mathbf{1}^{\top}\boldsymbol{\delta}=0$. Conditional on a locally fixed
quantization pattern, a differentiable reconstruction surrogate $J$ satisfies
\begin{equation}
\Delta J \approx \nabla J^{\top}\boldsymbol{\delta}
  +\tfrac{1}{2}\boldsymbol{\delta}^{\top}H\boldsymbol{\delta}.
\label{eq:mean_matched_taylor}
\end{equation}
Mean matching removes only displacement in the all-ones direction. It does not
remove alignment with layer-specific gradients or cross-layer curvature;
sequential prefix coupling can make $H$ non-diagonal, and permutation changes
that alignment. We did not estimate these derivatives, so
\xeqref{eq:mean_matched_taylor} is a local interpretation of the control,
not a Taylor model of WER or a mechanism test.

\subsection{Diagnostic ablation and layer behavior}
\label{sec:ablation_signals}
\label{sec:alpha_layers}

\xref{Table}{tab:ablation_signals} tests the gate terms at W3 on two models.
Either term improves on the constant midpoint; their combination has the
lowest mean and standard deviation. This conclusion is restricted to these
two runs.

\begin{center}
\begin{minipage}{\columnwidth}
\centering
\captionof{table}{Restricted diagnostic ablation at 3 bits on LibriSpeech
test-other. Values are mean WER $\pm$ standard deviation across five
calibration seeds. With both diagnostics disabled, the gate uses the interval
midpoint $\alpha=0.45$.}
\label{tab:ablation_signals}
\normalsize
\setlength{\tabcolsep}{4pt}
\begin{tabular}{ccrr}
\toprule
$\phi_{\rm int}$ & $\phi_{\rm sol}$ & W-Tiny & Q-0.6B \\
\midrule
\xmark & \xmark & 82.8 $\pm$ 15.4 & 47.8 $\pm$ 24.7 \\
\cmark & \xmark & 64.2 $\pm$ 16.2 & 37.6 $\pm$ 21.8 \\
\xmark & \cmark & 63.4 $\pm$ 16.0 & 38.4 $\pm$ 20.9 \\
\rowcolor{gray!8}
\cmark & \cmark & \textbf{62.3 $\pm$ 12.6} & \textbf{35.5 $\pm$ 20.0} \\
\bottomrule
\end{tabular}
\end{minipage}
\end{center}

The coefficient span is only 0.030--0.053, yet
\xrefrange{Appendix Figures}{fig:alpha_layers}{fig:alpha_histograms} show structured
layer and encoder--decoder variation. \xref{Figure}{fig:mechanism_compact}
summarizes structured reassignment;
\xref{Appendix Table}{tab:mechanism_controls} adds the dose results and exact values.
These aggregates do not identify causal mediation.

\subsection{FADE under a mixed-precision allocation}
\label{sec:adaptive_ptq}

The question is whether FADE's marginal effect survives a change of frozen
allocation host, not whether a stack outranks either component.
\adaptiveptqtable
\begin{figure}[t]
    \centering
    \includegraphics[width=\columnwidth]{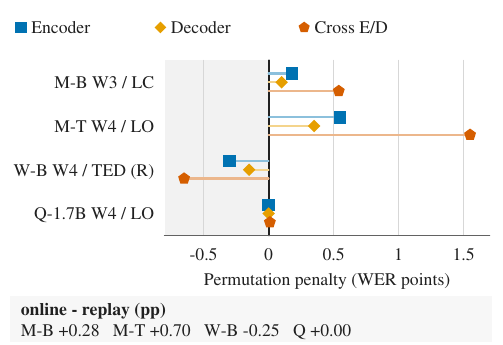}
    \caption{Structured reassignment on four diagnostic settings. Positive
    values mean that permutation raises WER relative to canonical FADE.
    Cross-boundary reassignment has the largest observed penalty in the two
    favorable rows; W-Base W4 TED reverses (R). The footer reports
    online-minus-replay in plot-row order. Exact values are in Appendix
    \xref{Appendix Table}{tab:mechanism_controls}.}
    \label{fig:mechanism_compact}
\end{figure}

The five unflagged favorable rows have negative $\Delta_F$ on both hosts. On
W4 TED, the effect changes from +2.3 on GPTQ to $-$0.3 on GenPTQ, so the
allocation offsets the harmful shape in this row. The high-variance row is
excluded and the near-tie is negligible.

\subsection{Offline cost and practical boundary}
\label{sec:cost}

\begin{figure}[t]
    \centering
    \includegraphics[width=\columnwidth]{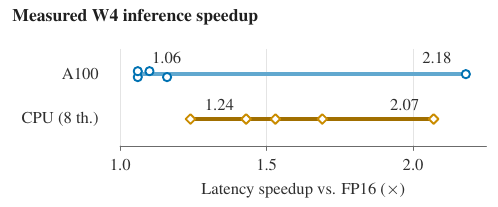}
    \caption{Our exported W4 models accelerate inference on both measured
    platforms. Each point denotes one model; endpoint labels give the observed
    range relative to FP16. Latency is the median of three timed runs after one
    warmup on an 11-second utterance. Appendix
    \xref{Appendix Table}{tab:inference_latency} reports the complete measurements.}
    \label{fig:deployment_efficiency}
\end{figure}

Our W4 exports give 1.06--2.18$\times$ A100 and 1.24--2.07$\times$ CPU
speedups (\xref{Figure}{fig:deployment_efficiency}).
\xref{Appendix}{app:efficiency} gives the protocol, offline cost, and per-model
latency, storage, and peak-memory measurements.

\subsection{Discussion}
\label{sec:discussion}

The result depends on comparison budget and bit regime. FADE is below QEP-0.5
in 31/38 settings without per-model search, but tuned \globalctrl recovers
much of the gap: 16/18/4 settings are lower/near/higher. Thirteen of 19 W3
settings favor FADE over \globalctrl; 15 of 19 W4 comparisons are near.
Five selected rows retain favorable paired and permutation evidence, whereas
one reverses and two remain unresolved. Thus the evidence supports assignment
sensitivity in particular settings, not a uniform layer-wise advantage.

Stable W4 rows often differ by hundredths of WER, while some W3 systems remain
too degraded for a relative gain to imply deployability. With a labeled
development set, \globalctrl is the first reference; without target-model
search, QEP-0.5 is budget matched. Absolute WER and seed variation remain part
of either deployment decision.

\section{Conclusion}
\label{sec:conclusion}

\methodname separates the model-level operating point from the residual
layer-wise pattern that a shared QEP coefficient conflates. Its two
weight-space diagnostics derive this pattern offline, without target-model
search or inference-graph changes. Tuned \globalctrl recovers much of
\methodname's gain over fixed QEP, so the operating point accounts for much of
the gap. Mean-matched and reassignment controls support placement effects in
selected settings.
Gains concentrate in degraded W3 rows, while a W4 reversal changes sign in the
GenPTQ-host contrast; the marginal effect varies across the displayed
host comparisons. \methodname provides a search-free way to use this
layer-wise pattern without development tuning, not a universal replacement for
tuned compensation.

\clearpage
\section*{Limitations}
This study covers weight-only post-training quantization for three
encoder--decoder ASR model families and four English test sets. It does not
establish the same behavior for other languages, decoder-only or transducer
architectures, activation quantization, or other runtime kernels. The claim of
avoiding per-model coefficient search concerns separate target-model
selection: the shared coefficient interval was chosen once on LibriSpeech
dev-other using three anchor models. By contrast,
\globalctrl and the Mean-Matched Shape Control use development-tuned
model--bit coefficients. The four non-anchor models are other sizes from the
same three families, and the two strata have different dataset coverage.
Their comparison supports reuse within the evaluated families, not
generalization to unseen model families or architectures.

The full matrix uses at least five calibration seeds, but paired confidence
intervals and within-seed permutation distributions are available only for an
eight-setting, ten-seed diagnostic subset. This subset was chosen after the
initial aggregate matrix to include five initially favorable cases, a
high-variance case,
a reversal, and a near-tie. It is therefore outcome-informed and cannot
estimate how often FADE improves over the full 38 settings. The remaining
control results are aggregate means and should not be read as significance
tests. The mixed-precision, mean-matched, and mechanism follow-ups are also
available only as aggregate summaries. They support descriptive contrasts,
not new paired inference.

The GenPTQ comparison matches average weight bits rather than serialized
bytes and lacks a GenPTQ+Global arm. Only the displayed within-allocation
contrasts are identified; cross-host WER levels and an
allocation-by-compensation interaction are not. Two nonzero dose levels do
not establish a scaling law. The diagnostic-component summary also lacks
per-layer records linking $g(l)$ to WER, so it measures activity frequency
rather than causal contribution. We did not run dedicated sensitivity studies
for the per-model group sizes or the additive fusion rule. Finally, latency is
measured on one 11-second utterance on an A100 and an eight-thread server CPU.
These measurements do not cover energy use, mobile hardware, or other runtime
backends.

\section*{Ethical Considerations}
This work studies model compression and does not introduce a new speech
dataset or collect personal data. All experiments use existing ASR models and
public benchmark datasets under their respective licenses. Compression can
make speech recognition easier to deploy on resource-constrained devices, but
the error patterns of the underlying models may persist or worsen after
quantization. Deployments should therefore evaluate performance across the
languages, accents, acoustic conditions, and user populations relevant to the
application. The paper reports aggregate word error rate and does not claim
that these averages establish demographic fairness.

\bibliography{cas-refs}

\clearpage
\appendix
\raggedbottom
\section{Offline algorithm and implementation protocol}
\label{app:algorithm}

\xref{Algorithm}{alg:fade} gives the complete offline procedure. The RTN and
calibrated solutions use the same bit width and group size; the calibrated
solver selects its scale and clipping multiplier as specified below.

\begin{algorithm}[h!]
\caption{\methodname offline quantization}
\label{alg:fade}
\begin{algorithmic}[1]
\REQUIRE Weights $\{\mat W_l\}_{l=1}^L$, calibration data, bit width $b$,
bounds $\alpha_{\min},\alpha_{\max}$
\ENSURE Quantized weights $\{\widehat{\mat W}_l\}_{l=1}^L$
\FOR{$l=1$ to $L$}
  \STATE Collect clean $\mat X_l$ and quantized-prefix $\widehat{\mat X}_l$
  \STATE $\widehat{\mat W}^{\rm rtn}_l\leftarrow
  \mathrm{RTN}(\mat W_l,b)$
  \STATE $\widehat{\mat W}^{\rm cal}_l\leftarrow
  \mathrm{GPTQ}(\mat W_l,\widehat{\mat X}_l,b)$
  \STATE Compute $e_r(l)$, $e_c(l)$, $g(l)$, and $d(l)$
  \STATE Compute $\phi_{\rm int}(l)$ and $\phi_{\rm sol}(l)$
  \STATE Set $\alpha_l$ with \xeqref{eq:fusion_gate}
  \STATE Solve \xeqref{eq:unified_objective} for $\widehat{\mat W}_l$
\ENDFOR
\RETURN $\{\widehat{\mat W}_l\}_{l=1}^L$
\end{algorithmic}
\end{algorithm}

\paragraph{Quantized matrices and solver.}
We quantize every linear projection in encoder and decoder attention and
feed-forward blocks. Embeddings, normalization parameters, convolutional
front ends, and output heads remain in FP16. For activation matrix $X$, GPTQ
uses
$H=XX^{\top}/N+0.01\,\operatorname{mean}(\operatorname{diag}(XX^{\top}/N))I$.
Columns are processed in descending Hessian-diagonal order, in blocks of 128,
with the standard sequential error update. The calibrated and corrected solves
reuse the same Cholesky factorization.

\paragraph{Scale and clipping.}
Each output-row group uses unsigned codes in $[0,2^b-1]$ and an asymmetric
scale and zero point. RTN uses the group's observed minimum and maximum. For
GPTQ and the corrected solve, we test clipping multipliers
$\rho\in\{1.00,0.99,\ldots,0.21\}$, recompute the scale and zero point for
each $\rho$, and retain the grid with the lowest groupwise squared weight
error before the Hessian reconstruction. No outlier channel is left in FP16.
The solver-selected grid is also the grid used to compute $e_c(l)$ and $d(l)$.

\paragraph{Calibration, decoding, and WER.}
Seeds 0--4 sample 128 train-clean-100 utterances without replacement; the
ten-seed subset uses seeds 0--9. Audio is resampled to 16~kHz and passed
through each checkpoint's official feature processor without stochastic
augmentation. Decoder calibration uses teacher forcing on the reference
transcript, conditioned on the current quantized encoder prefix. Evaluation
uses greedy autoregressive decoding (one beam, no sampling or external
language model) with the English transcription prompt where the checkpoint
requires one. References and hypotheses are Unicode-normalized, lowercased,
stripped of punctuation, and whitespace-collapsed before corpus WER is
computed from total substitutions, deletions, and insertions.

\FloatBarrier
\section{Statistical and assignment controls}
\label{app:controls}

\providecommand{\anchorsplittable}{%
\centering
\scriptsize
\textbf{(a) Anchor versus non-anchor outcomes}\\[3pt]
\setlength{\tabcolsep}{2.2pt}
\begin{tabular}{@{}lcccc@{}}
\toprule
& \multicolumn{2}{c}{Exact L/E/H} & \multicolumn{2}{c}{Tolerance L/N/H} \\
\cmidrule(lr){2-3}\cmidrule(lr){4-5}
Control & Anchor & Non-anchor & Anchor & Non-anchor \\
\midrule
QEP-0.5 & 12/0/0 & 19/0/7 & 8/4/0 & 15/7/4 \\
\globalctrl & 8/3/1 & 13/1/12 & 8/4/0 & 8/14/4 \\
Mean & 12/0/0 & 14/0/12 & 8/4/0 & 9/13/4 \\
E/D & 8/3/1 & 15/2/9 & 7/5/0 & 6/17/3 \\
Perm & 11/0/1 & 18/2/6 & 8/4/0 & 9/15/2 \\
\bottomrule
\end{tabular}

\vspace{7pt}
\textbf{Non-anchor breakdown by bit width}\\[3pt]
\setlength{\tabcolsep}{4.4pt}
\begin{tabular}{@{}llrcc@{}}
\toprule
Control & Bits & $N$ & Exact L/E/H & Tolerance L/N/H \\
\midrule
QEP-0.5 & W3 & 13 & 9/0/4 & 8/2/3 \\
QEP-0.5 & W4 & 13 & 10/0/3 & 7/5/1 \\
\globalctrl & W3 & 13 & 8/0/5 & 7/3/3 \\
\globalctrl & W4 & 13 & 5/1/7 & 1/11/1 \\
\bottomrule
\end{tabular}
}

\providecommand{\protocolstatstable}{%
\centering
\scriptsize
\textbf{(b) Selected \globalctrl and permutation summary}\\[3pt]
\setlength{\tabcolsep}{3pt}
\renewcommand{\arraystretch}{0.98}
\begin{tabular}{lrr}
\toprule
\multicolumn{3}{c}{Selected \globalctrl} \\
Model & W3 & W4 \\
\midrule
Whisper-Tiny & 0.40 & 0.45 \\
Whisper-Base & 0.45 & 0.45 \\
Whisper-Small & 0.50 & 0.45 \\
Moonshine-Tiny & 0.40 & 0.40 \\
Moonshine-Base & 0.45 & 0.45 \\
Qwen-0.6B & 0.40 & 0.45 \\
Qwen-1.7B & 0.40 & 0.45 \\
\midrule
\multicolumn{3}{c}{Within-seed layer permutation} \\
Setting & Median [5th, 95th] & \methodname pct. \\
\midrule
W-Tiny W4 LO & +1.10 [+0.30, +2.00] & $\sim$4\% \\
M-Tiny W4 LO & +2.10 [+0.50, +4.20] & $\sim$2\% \\
Q-1.7B W3 TED & +1.50 [+0.20, +3.40] & $\sim$6\% \\
W-Base W3 TED & +1.30 [+0.10, +2.80] & $\sim$8\% \\
M-Base W3 LC & +0.62 [+0.35, +0.95] & $\sim$1\% \\
Q-0.6B W3 SPG & +2.00 [$-$5.00, +12.00] & $\sim$25\% \\
W-Base W4 TED & $-$0.60 [$-$3.00, +1.50] & $\sim$70--85\% \\
Q-1.7B W4 LO & $-$0.01 [$-$0.04, +0.03] & $\sim$50--65\% \\
\bottomrule
\end{tabular}
}

\begin{table*}[t]
\centering
\caption{Selection-stratified outcomes and assignment controls. Panel (a)
separates the three interval-selection anchors (12 settings) from the other
    four model sizes (26 settings); notation follows
    \xref{Table}{tab:control_summary}. Panel (b) reports the development-selected
\globalctrl coefficients and pools 100 layer reassignments per seed over ten
seeds. For permutation, positive
$\Delta=\mathrm{WER}_{\rm Perm}-\mathrm{WER}_{\rm FADE}$ favors the canonical
assignment.}
\label{tab:control_anchor_split}
\label{tab:protocol_statistics}
\begin{minipage}[t]{0.51\textwidth}
\vspace{0pt}
\anchorsplittable
\end{minipage}
\hfill
\begin{minipage}[t]{0.46\textwidth}
\vspace{0pt}
\protocolstatstable
\end{minipage}
\end{table*}

\globalctrl is selected on LibriSpeech dev-other using five calibration seeds
and one grid value per model--bit pair, then applied unchanged to LC, LO, SPG,
and TED. Mean, E/D, and Perm are reconstructed within each calibration seed
using the same utterance IDs as \methodname. Perm reassigns coefficients
within the seed and reruns corrected quantization.
The eight-setting diagnostic subset was frozen before its ten-seed rerun. It
contains five initially favorable cases, one high-variance case, one reversal, and one
near-tie, so it is designed for diagnosis rather than prevalence estimation.

\paragraph{Development selection.}
For an interval candidate $c$, each anchor-model--bit--seed dev WER is divided
by the minimum over the six candidates for that same cell; \xref{Table}{tab:interval_selection}
reports the arithmetic mean of these 30 ratios. For \globalctrl, we average
dev-other WER over seeds 0--4 at each grid value. An exact tie is resolved by
choosing the coefficient closest to 0.5, then the smaller coefficient.

\newpage

\paragraph{Paired intervals and permutations.}
For each diagnostic setting, let
$d_s=\mathrm{WER}_{\rm Global,s}-\mathrm{WER}_{\rm FADE,s}$. The reported
95\% interval is the paired Student interval
$\bar d\pm t_{0.975,9}s_d/\sqrt{10}$, with sample standard deviation
($\mathrm{ddof}=1$); no multiplicity correction is applied. For each
calibration seed, we draw 100 unrestricted permutations of the complete
encoder--decoder coefficient vector, preserve its multiset, and rerun
sequential corrected quantization from the FP16 checkpoint. Permutation $j$
under calibration seed $s$ uses random seed $1000s+j$.
\xref{Table}{tab:protocol_statistics} pools the resulting 1,000 seed-matched
$\mathrm{WER}_{\rm Perm}-\mathrm{WER}_{\rm FADE}$ differences per setting.
The full-matrix Perm mean first averages the 100 permutations within each seed
and then averages over seeds.

\xref{Tables}{tab:control_summary} and~\numref{tab:control_full} report the complete
38-setting control matrix. The counts are descriptive because paired records
are available only for the frozen diagnostic subset.

\providecommand{\controlsummarytable}{%
\begin{table}[!t]
\centering
\caption{WER outcomes for \methodname against QEP-0.5, development-tuned
\globalctrl, and assignment controls over all 38 matched settings. Exact
L/E/H counts settings where
\methodname WER is lower, equal, or higher at the reported precision. The
tolerance column uses
$\tau=\max(0.05,0.01\,\mathrm{WER}_{\textsc{Fade}})$ and counts lower, near,
and higher. Both summaries treat settings as comparison units and are
descriptive, not significance tests.}
\label{tab:control_summary}
\normalsize
\setlength{\tabcolsep}{4pt}
\begin{tabular}{lcc}
\toprule
Control & Exact L/E/H & Tolerance L/N/H \\
\midrule
QEP-0.5 & 31/0/7 & 23/11/4 \\
\globalctrl & 21/4/13 & 16/18/4 \\
Mean & 26/0/12 & 17/17/4 \\
E/D & 23/5/10 & 13/22/3 \\
Perm & 29/2/7 & 17/19/2 \\
\bottomrule
\end{tabular}
\end{table}
}

\controlsummarytable

\xref{Table}{tab:control_anchor_split} shows that the non-anchor comparison
stays within the same Whisper, Moonshine, and Qwen3-ASR families and is not an
unseen-architecture test. At non-anchor W4, 11 of 13 comparisons with
\globalctrl are near under the descriptive tolerance. In the same five
favorable diagnostic rows, \methodname falls in the best 1--8\% of
permutations; the high-variance, reversal, and near-tie cases do not follow
that pattern.

\begin{table*}[!t]
\centering
\caption{Aggregate WER ($\downarrow$) for all 38 coefficient-control
settings. Bold marks the lowest displayed value within each
model--dataset--bit setting. These aggregates are descriptive rather than
paired significance tests.}
\label{tab:control_full}
\scriptsize
\renewcommand{\arraystretch}{1.35}
\begin{minipage}[t]{0.495\textwidth}
\centering
\vspace{0pt}
\setlength{\tabcolsep}{1.2pt}
\begin{tabular}{llrrrrrr}
\toprule
Model & Setting & QEP-0.5 & \globalctrl & Mean & E/D & Perm & \methodname \\
\midrule
\multicolumn{8}{c}{Whisper} \\
\midrule
Tiny & W3 LO & 94.62 & 73.62 & 78.46 & 68.13 & 75.23 & \textbf{62.31} \\
 & W4 LO & 32.77 & 30.89 & 31.33 & 30.40 & 31.04 & \textbf{29.88} \\
Base & W3 LO & 33.43 & \textbf{33.33} & 33.60 & 33.73 & 33.83 & 34.10 \\
 & W4 LO & 16.92 & 16.46 & 16.57 & 16.34 & 16.49 & \textbf{16.21} \\
 & W3 LC & 16.45 & 15.00 & 15.34 & 14.62 & 15.11 & \textbf{14.22} \\
 & W4 LC & \textbf{4.86} & \textbf{4.86} & \textbf{4.86} & \textbf{4.86} & 4.87 & 4.87 \\
 & W3 SPG & 26.90 & \textbf{26.88} & 26.93 & 26.95 & 26.97 & 27.02 \\
 & W4 SPG & 15.56 & 15.40 & 15.44 & 15.37 & 15.42 & \textbf{15.34} \\
 & W3 TED & 35.59 & 33.34 & 33.86 & 32.75 & 33.51 & \textbf{32.13} \\
 & W4 TED & 20.93 & \textbf{20.73} & 21.44 & 21.86 & 22.17 & 22.99 \\
Small & W3 LO & 10.79 & \textbf{10.76} & 10.83 & 10.87 & 10.89 & 10.96 \\
 & W4 LO & 11.33 & \textbf{11.15} & 11.16 & 11.19 & 11.23 & 11.17 \\
\midrule
\multicolumn{8}{c}{Moonshine} \\
\midrule
Tiny & W3 LO & 298.53 & 223.26 & 240.63 & 203.57 & 229.05 & \textbf{182.73} \\
 & W4 LO & 21.83 & 18.19 & 19.03 & 17.24 & 18.47 & \textbf{16.23} \\
Base & W3 LO & 11.74 & \textbf{11.61} & 11.62 & 11.62 & 11.67 & 11.63 \\
 & W4 LO & \textbf{9.39} & \textbf{9.39} & \textbf{9.39} & \textbf{9.39} & 9.40 & 9.40 \\
 & W3 LC & 4.73 & 3.67 & 3.92 & 3.39 & 3.75 & \textbf{3.10} \\
 & W4 LC & 4.08 & \textbf{3.85} & \textbf{3.85} & 3.89 & 3.94 & 3.86 \\
 & W3 SPG & 9.28 & 9.12 & 9.16 & 9.09 & 9.14 & \textbf{9.06} \\
 & W4 SPG & 8.13 & \textbf{8.08} & \textbf{8.08} & \textbf{8.08} & \textbf{8.08} & 8.09 \\
 & W3 TED & 18.19 & 17.73 & 17.84 & 17.61 & 17.76 & \textbf{17.48} \\
 & W4 TED & 16.52 & 16.41 & 16.44 & 16.39 & 16.42 & \textbf{16.37} \\
\bottomrule
\end{tabular}
\end{minipage}
\hfill
\begin{minipage}[t]{0.495\textwidth}
\centering
\vspace{0pt}
\setlength{\tabcolsep}{1.2pt}
\begin{tabular}{llrrrrrr}
\toprule
Model & Setting & QEP-0.5 & \globalctrl & Mean & E/D & Perm & \methodname \\
\midrule
\multicolumn{8}{c}{Qwen3-ASR} \\
\midrule
0.6B & W3 LC & 30.23 & \textbf{30.05} & 30.54 & 30.78 & 30.96 & 31.45 \\
 & W3 LO & 37.93 & 36.32 & 36.69 & 35.90 & 36.44 & \textbf{35.45} \\
 & W3 SPG & 46.25 & 42.01 & 42.98 & 40.90 & 42.33 & \textbf{39.72} \\
 & W3 TED & 48.14 & 44.17 & 45.08 & 43.13 & 44.47 & \textbf{42.03} \\
 & W4 LC & 2.36 & \textbf{2.32} & 2.33 & \textbf{2.32} & 2.33 & \textbf{2.32} \\
 & W4 LO & 5.03 & \textbf{4.96} & \textbf{4.96} & 4.97 & 4.98 & 4.97 \\
 & W4 SPG & 3.25 & 3.16 & 3.18 & 3.15 & 3.17 & \textbf{3.13} \\
 & W4 TED & 10.30 & 10.20 & 10.22 & 10.18 & 10.21 & \textbf{10.16} \\
1.7B & W3 LC & 2.89 & 2.53 & 2.62 & 2.44 & 2.56 & \textbf{2.34} \\
 & W3 LO & 5.86 & 5.48 & 5.56 & 5.38 & 5.51 & \textbf{5.27} \\
 & W3 SPG & 5.66 & 5.40 & 5.47 & 5.37 & 5.44 & \textbf{5.32} \\
 & W3 TED & 15.63 & 13.03 & 13.63 & 12.35 & 13.23 & \textbf{11.63} \\
 & W4 LC & 1.72 & \textbf{1.70} & 1.71 & \textbf{1.70} & 1.71 & \textbf{1.70} \\
 & W4 LO & 3.62 & \textbf{3.58} & 3.60 & \textbf{3.58} & \textbf{3.58} & 3.59 \\
 & W4 SPG & 2.89 & \textbf{2.88} & 2.89 & \textbf{2.88} & 2.89 & \textbf{2.88} \\
 & W4 TED & 10.15 & \textbf{10.07} & 10.08 & \textbf{10.07} & 10.08 & \textbf{10.07} \\
\bottomrule
\end{tabular}
\end{minipage}
\end{table*}

\xref{Figure}{fig:core_diagnostic_evidence} condenses the paired and
mean-matched evidence used in the main text. The following tables retain the
complete numerical values.

\providecommand{\coreevidencetable}{%
\begin{table*}[!t]
\centering
\caption{Ten-seed evidence on the frozen diagnostic subset. WER values are
mean $\pm$ standard deviation. The paired Student 95\% CI is for
WER$_{\rm Global}-$WER$_{\rm FADE}$, so positive intervals favor FADE.
Perm. p5 is the fifth percentile of WER$_{\rm Perm}-$WER$_{\rm FADE}$;
``pct.'' is FADE's percentile in the within-seed permutation distribution.}
\label{tab:core_evidence_summary}
\scriptsize
\setlength{\tabcolsep}{1.8pt}
\renewcommand{\arraystretch}{1.01}
\begin{tabular*}{\textwidth}{@{\extracolsep{\fill}}lrrrrrr@{}}
\toprule
Setting & Global & E/D & FADE & Global$-$FADE CI & Perm. p5 & FADE pct. \\
\midrule
W-Tiny W4 LO & 30.9$\pm$1.5 & 30.4$\pm$1.3 & 29.9$\pm$1.0 & [0.58, 1.44] & +0.30 & 4\% \\
M-Tiny W4 LO & 18.2$\pm$2.2 & 17.2$\pm$1.7 & 16.2$\pm$1.0 & [1.28, 2.64] & +0.50 & 2\% \\
Q-1.7B W3 TED & 13.0$\pm$5.0 & 12.4$\pm$4.5 & 11.6$\pm$4.2 & [0.58, 2.22] & +0.20 & 6\% \\
W-Base W3 TED & 33.3$\pm$3.2 & 32.8$\pm$2.8 & 32.1$\pm$2.5 & [0.46, 1.96] & +0.10 & 8\% \\
M-Base W3 LC & 3.67$\pm$0.25 & 3.39$\pm$0.22 & 3.10$\pm$0.19 & [0.47, 0.67] & +0.35 & 1\% \\
Q-0.6B W3 SPG$^{\dagger}$ & 42.0$\pm$34 & 40.9$\pm$33 & 39.7$\pm$34 & [$-$3.07, 7.65] & $-$5.00 & 25\% \\
W-Base W4 TED$^{\ddagger}$ & 20.7$\pm$2.7 & 21.9$\pm$3.5 & 23.0$\pm$4.4 & [$-$3.55, $-$0.97] & $-$3.00 & 70--85\% \\
Q-1.7B W4 LO & 3.58$\pm$0.02 & 3.58$\pm$0.02 & 3.59$\pm$0.01 & [$-$0.02, 0.00] & $-$0.04 & 50--65\% \\
\bottomrule
\end{tabular*}
\vspace{1pt}
\begin{minipage}{0.98\textwidth}
\scriptsize
$^{\dagger}$High-variance row, excluded from directional claims.
$^{\ddagger}$Reversal selected deliberately to represent an adverse case.
\end{minipage}
\end{table*}
}

\coreevidencetable

\centereddecompositiontable

\xref{Table}{tab:core_statistics} reports the ten-seed paired rerun. Five
intervals favor \methodname, one favors \globalctrl, and two include zero.

\begin{table*}[!t]
\centering
\caption{Ten-seed reruns on the frozen diagnostic subset, with shared
calibration utterances. Values are mean WER $\pm$ standard deviation. The
final column gives the paired Student 95\% confidence interval for \globalctrl minus
\methodname; a positive interval favors \methodname. Bold marks the lowest
displayed mean in each row.}
\label{tab:core_statistics}
\small
\setlength{\tabcolsep}{2.4pt}
\renewcommand{\arraystretch}{1.30}
\begin{tabular*}{\textwidth}{@{\extracolsep{\fill}}lrrrrrr@{}}
\toprule
Setting & \globalctrl & Mean & E/D & Perm & \methodname &
\makecell{\globalctrl{} $-$ \methodname\\95\% CI} \\
\midrule
W-Tiny W4 LO & 30.9$\pm$1.5 & 31.3$\pm$1.7 & 30.4$\pm$1.3 &
31.0$\pm$1.5 & \textbf{29.9$\pm$1.0} & [0.58, 1.44] \\
M-Tiny W4 LO & 18.2$\pm$2.2 & 19.0$\pm$2.6 & 17.2$\pm$1.7 &
18.5$\pm$2.1 & \textbf{16.2$\pm$1.0} & [1.28, 2.64] \\
Q-1.7B W3 TED & 13.0$\pm$5.0 & 13.6$\pm$5.2 & 12.4$\pm$4.5 &
13.2$\pm$4.8 & \textbf{11.6$\pm$4.2} & [0.58, 2.22] \\
W-Base W3 TED & 33.3$\pm$3.2 & 33.9$\pm$3.4 & 32.8$\pm$2.8 &
33.5$\pm$3.1 & \textbf{32.1$\pm$2.5} & [0.46, 1.96] \\
M-Base W3 LC & 3.67$\pm$0.25 & 3.92$\pm$0.30 & 3.39$\pm$0.22 &
3.75$\pm$0.28 & \textbf{3.10$\pm$0.19} & [0.47, 0.67] \\
\midrule
Q-0.6B W3 SPG & 42.0$\pm$34 & 43.0$\pm$34 & 40.9$\pm$33 &
42.3$\pm$34 & \textbf{39.7$\pm$34} & [$-$3.07, 7.65] \\
W-Base W4 TED & \textbf{20.7$\pm$2.7} & 21.4$\pm$3.0 & 21.9$\pm$3.5 &
22.2$\pm$4.0 & 23.0$\pm$4.4 & [$-$3.55, $-$0.97] \\
Q-1.7B W4 LO & \textbf{3.58$\pm$0.02} & 3.60$\pm$0.02 &
\textbf{3.58$\pm$0.02} & \textbf{3.58$\pm$0.02} & 3.59$\pm$0.01 &
[$-$0.02, 0.00] \\
\bottomrule
\end{tabular*}
\end{table*}

\xref{Table}{tab:mechanism_controls} reports contrasts for coefficient amplitude,
within-module assignment, cross-boundary assignment, and sequential replay.
These aggregate controls do not identify causal mediation.

\mechanismcontrolstable

\section{Extended established-baseline results}
\label{app:extended_baselines}

\xref{Tables}{tab:cross_dataset} and~\numref{tab:qwen_full} extend the
established-baseline comparison beyond LibriSpeech test-other. The former
reports cross-domain results for Whisper-Base and Moonshine-Base; the latter
reports both Qwen3-ASR sizes on all four test sets.

\coveragefamilytable

Relative to QEP-0.5, \methodname lowers mean WER in 7/12 Whisper, 9/10
Moonshine, and 15/16 Qwen3-ASR settings. The exceptions include 3-bit
Whisper-Base on SPGISpeech, where QEP is slightly lower, and 4-bit
Whisper-Base on TED-LIUM~3, where RTN is lower. These comparisons measure
transfer across English speech domains, not multilingual generalization.

For Qwen3-ASR-1.7B, \methodname is lower than QEP-0.5 on all four W3 test
sets. The 0.6B W3 runs are degraded and highly variable. At W4, the absolute
QEP--\methodname differences range from 0.01 to 0.14 WER across the two model
sizes.

\crossdomaintable
\qwenfulltable

\begin{table*}[!t]
    \centering
    \includegraphics[width=\textwidth]{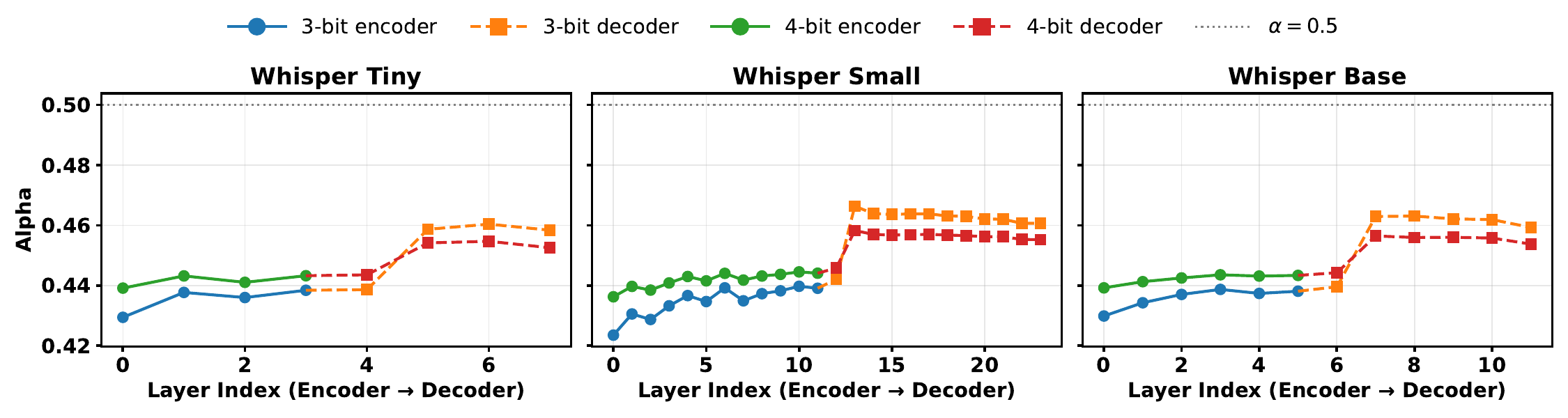}
    \captionof{figure}{Layer-wise FADE coefficients for W3 and W4 Whisper quantization
    across three model sizes. Each panel concatenates the encoder and decoder;
    the vertical transition marks their boundary.}
    \label{fig:alpha_layers}
\end{table*}

\FloatBarrier

\section{Coefficient and sensitivity analyses}
\label{app:diagnostic_figures}

Calibration uses 128 utterances from LibriSpeech train-clean-100 per seed. The
coefficient distributions and sensitivity plots use the interval selected on
dev-other with five seeds and locked before test evaluation. The aggregate
follow-up records omit layer-level links between diagnostics and WER, so the
figures describe coefficient structure rather than diagnostic causality.

\xref{Figure}{fig:alpha_layers} shows the coefficient trajectory used by FADE.
Its narrow range is not constant: encoder and decoder blocks follow different
paths, and the W3 and W4 paths separate within each model.

\paragraph{Why these bounds?}
The coefficient has a natural non-extrapolating domain: $\alpha_l=0$ recovers
the uncorrected reconstruction objective, while $\alpha_l=1$ applies the full
QEP target shift in \xeqref{eq:unified_objective}. We therefore restricted
interval candidates to this domain and compared six coarse choices using only
anchor-model dev-other WER. \xref{Table}{tab:interval_selection} shows that
$[0.1,0.8]$ obtained the lowest normalized score (1.000 versus 1.006--1.028),
after which it was frozen. On the eight-setting diagnostic subset, realized
coefficients occupy only 0.418--0.471
(\xref{Table}{tab:centered_decomposition}), so the endpoints bound the gate
rather than describe its typical output. We treat them as shared
development-selected bounds, not theoretically optimal values.

\xref{Figure}{fig:legacy_diagnostics} includes a fixed-coefficient sweep on a
separate 100-example subset with three runs; it is exploratory and does not
select a test-set configuration.

\intervalselectiontable

\begin{center}
\begin{minipage}{\columnwidth}
\centering
\captionof{table}{Summary of the eight-setting diagnostic-component rerun.
The aggregate records give ranges and do not distinguish strict wins from
ties.}
\label{tab:diagnostic_summary}
\small
\setlength{\tabcolsep}{5pt}
\begin{tabular}{lr}
\toprule
Outcome & Observed result \\
\midrule
Full gate best or tied best & 5--6 / 8 \\
W3 gain over best single term & 0.3--2.0 WER \\
W4 gain over best single term & 0--0.3 WER \\
Layers with $g(l)>0$ & 0--15\% \\
\bottomrule
\end{tabular}
\end{minipage}
\end{center}

The benefit of combining the two diagnostic terms is concentrated at W3: the
gain over the better single-term variant is 0.3--2.0 WER at W3 and 0--0.3 WER
at W4. The positive $g(l)$ term is active in at most 15\% of layers, but
activation frequency alone does not identify its contribution to WER.
\xref{Main-text Table}{tab:ablation_signals} gives the exact two-model ablation.

\begin{figure*}[t]
    \centering
    \begin{subfigure}[t]{0.48\textwidth}
        \centering
        \includegraphics[width=\linewidth]{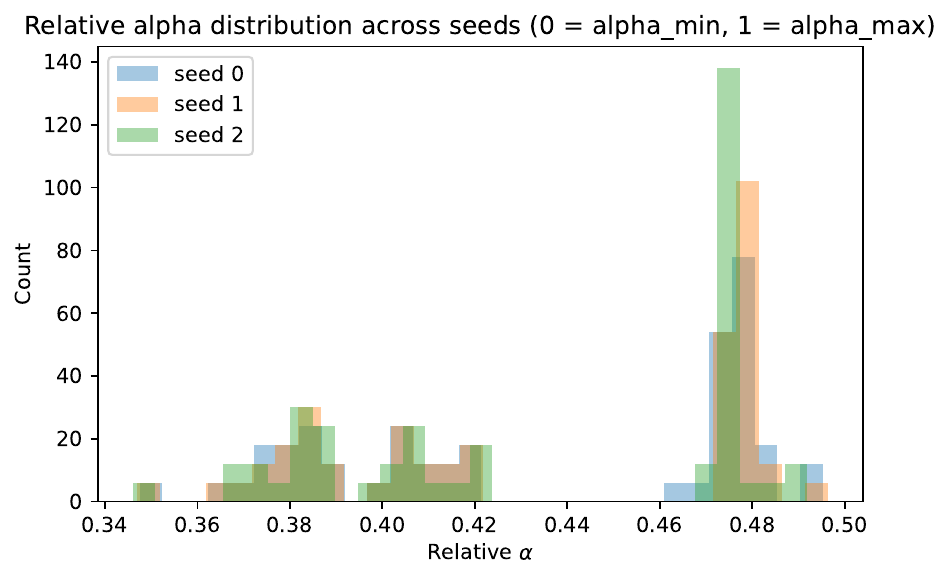}
        \caption{3-bit coefficients.}
        \label{fig:alpha_hist_3bit}
    \end{subfigure}
    \hfill
    \begin{subfigure}[t]{0.48\textwidth}
        \centering
        \includegraphics[width=\linewidth]{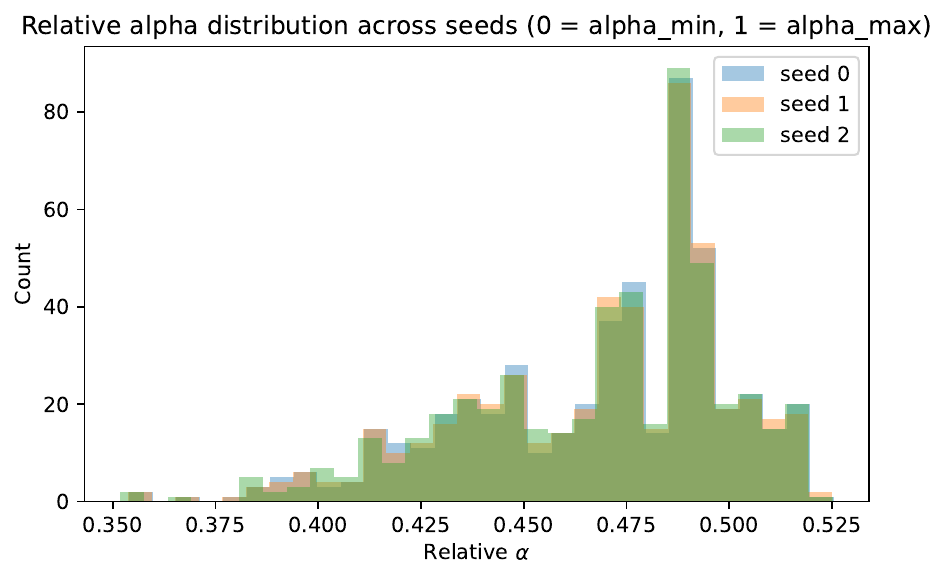}
        \caption{4-bit coefficients.}
        \label{fig:alpha_hist_4bit}
    \end{subfigure}
    \caption{Whisper-Tiny coefficient distributions across three calibration
    seeds at W3 and W4. Values are normalized as
    $(\alpha_l-\alpha_{\min})/(\alpha_{\max}-\alpha_{\min})$; the panels show
    similar marginal distributions across seeds.}
    \label{fig:alpha_histograms}
\end{figure*}

\begin{figure*}[t]
    \centering
    \begin{subfigure}[t]{0.40\textwidth}
        \centering
        \includegraphics[width=\linewidth]{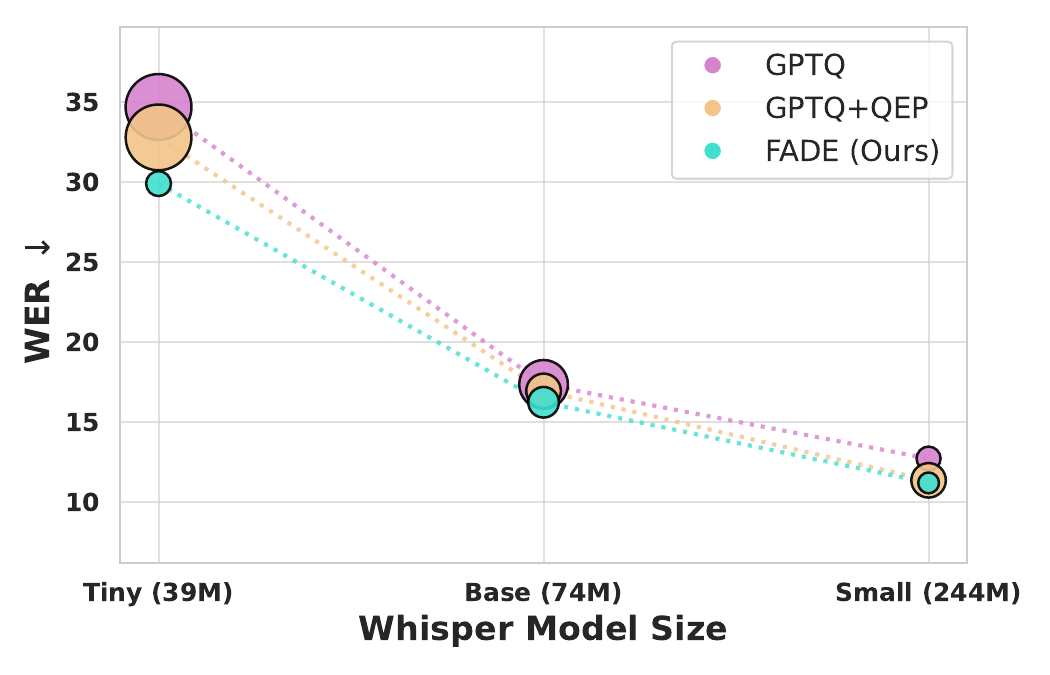}
        \caption{Whisper W4 established baselines.}
        \label{fig:legacy_whisper}
    \end{subfigure}
    \hfill
    \begin{subfigure}[t]{0.56\textwidth}
        \centering
        \includegraphics[width=\linewidth]{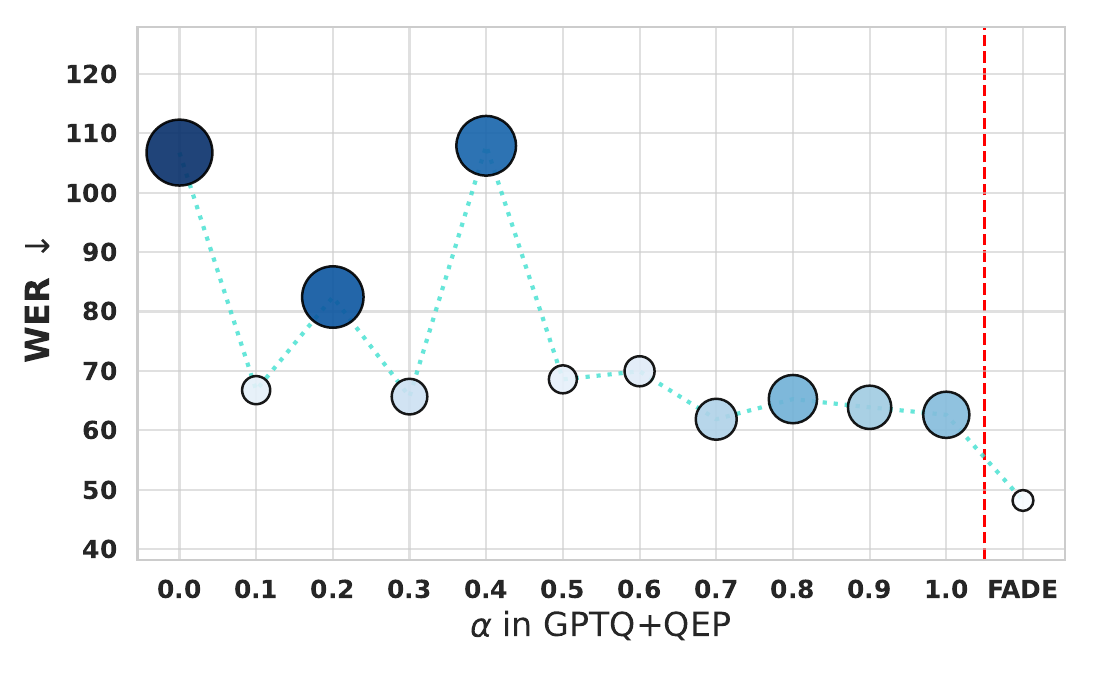}
        \caption{Whisper-Tiny W3 fixed-coefficient sweep.}
        \label{fig:alpha_sweep}
    \end{subfigure}
    \caption{Whisper coefficient diagnostics. Panel (a) compares
    \methodname, GPTQ, and QEP-0.5 at W4 on LibriSpeech test-other. Panel (b)
    evaluates fixed coefficients at W3 on a separate 100-example subset.
    Bubble area encodes the standard deviation across five calibration seeds
    in panel (a) and three runs in panel (b). The main analysis instead
    compares \methodname with the development-tuned \globalctrl control on the
    full test sets.}
    \label{fig:legacy_diagnostics}
\end{figure*}

\begin{figure*}[t]
    \centering
    \begin{subfigure}[t]{0.48\textwidth}
        \centering
        \includegraphics[width=\linewidth]{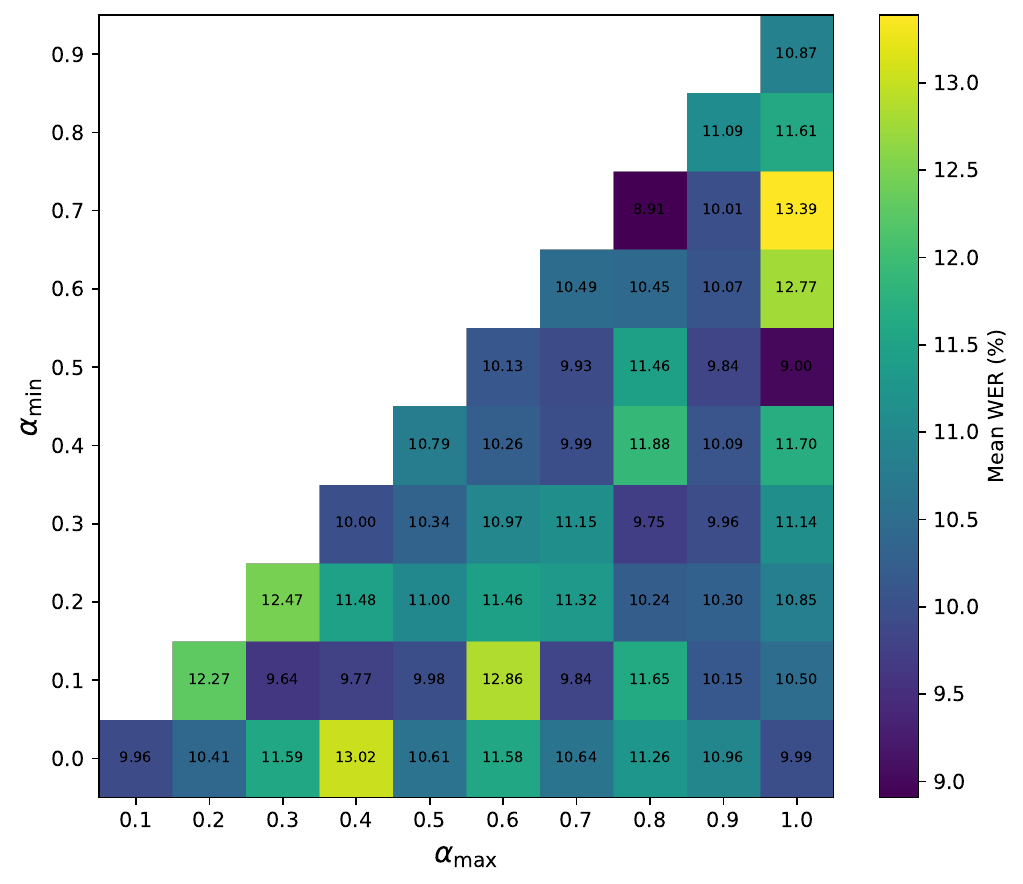}
        \caption{4-bit mean WER}
    \end{subfigure}
    \hfill
    \begin{subfigure}[t]{0.48\textwidth}
        \centering
        \includegraphics[width=\linewidth]{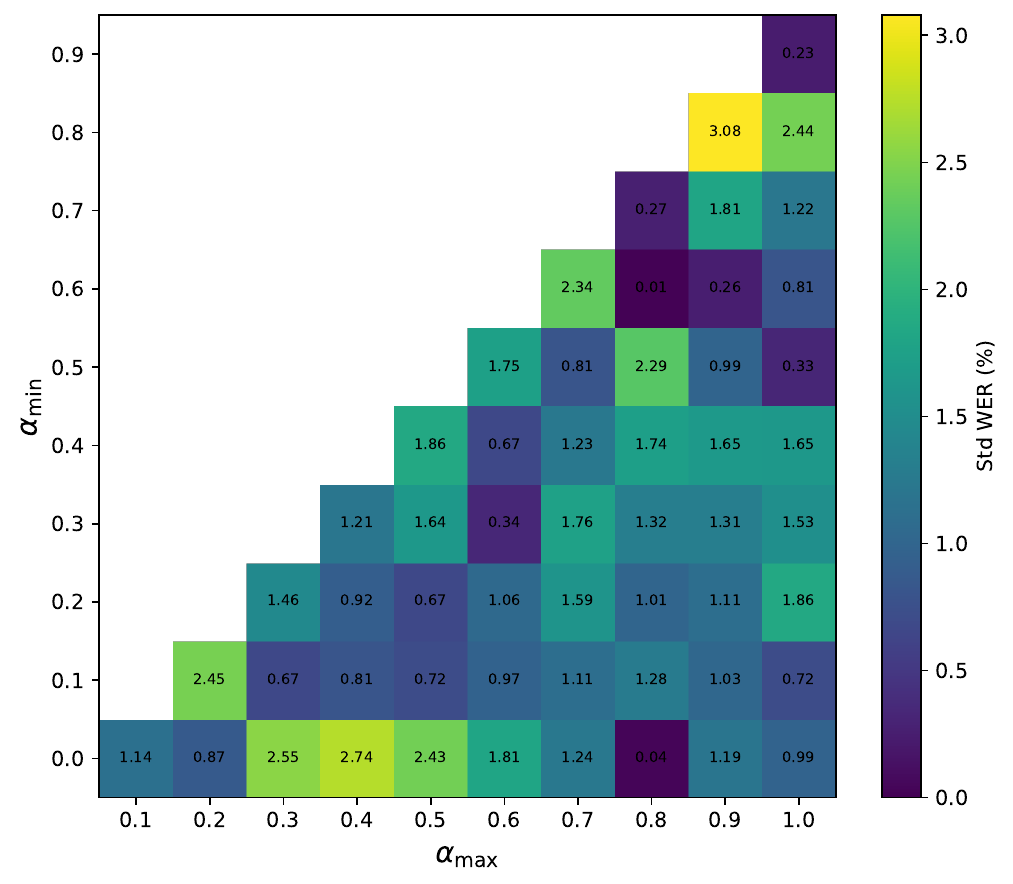}
        \caption{4-bit standard deviation}
    \end{subfigure}

    \begin{subfigure}[t]{0.48\textwidth}
        \centering
        \includegraphics[width=\linewidth]{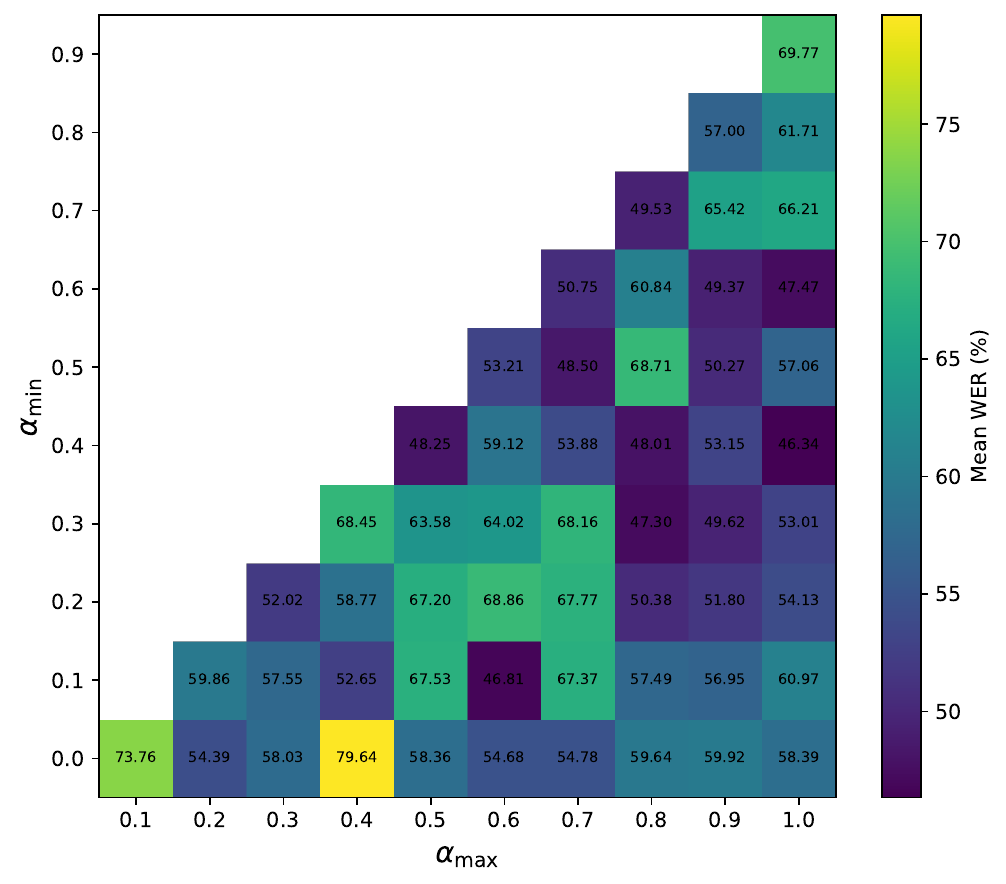}
        \caption{3-bit mean WER}
    \end{subfigure}
    \hfill
    \begin{subfigure}[t]{0.48\textwidth}
        \centering
        \includegraphics[width=\linewidth]{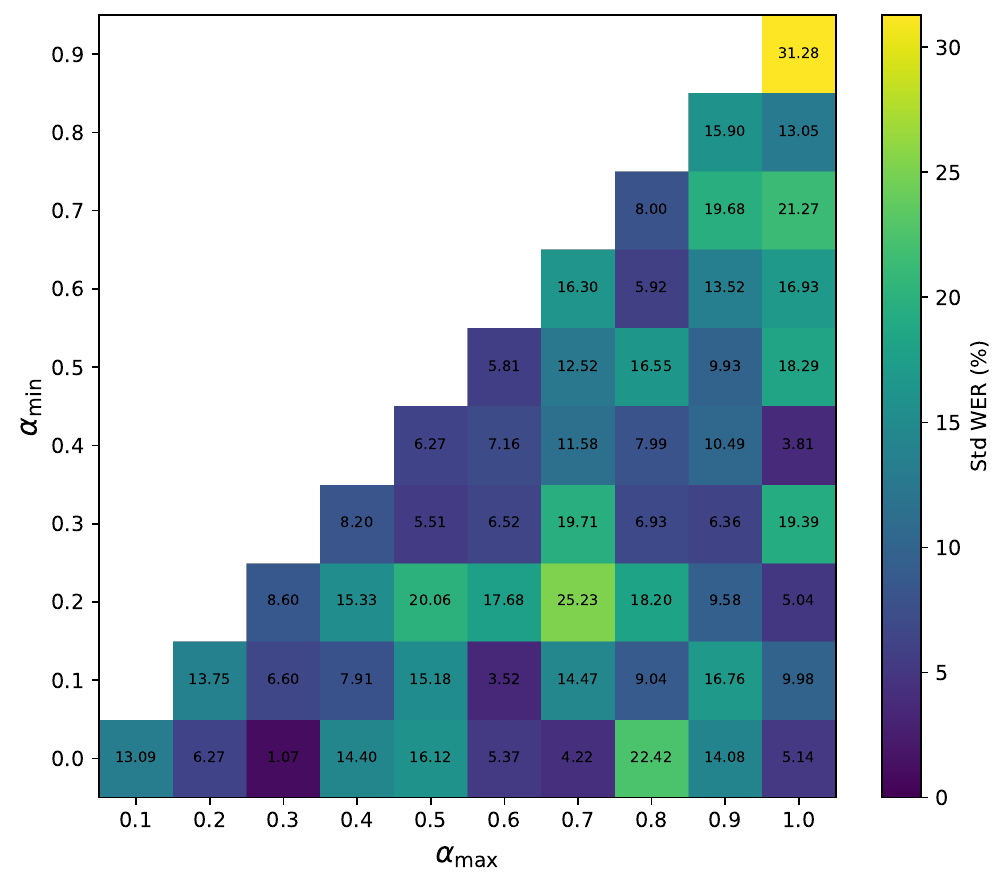}
        \caption{3-bit standard deviation}
    \end{subfigure}
    \caption{Sensitivity of Whisper-Tiny to the coefficient interval
    $(\alpha_{\min},\alpha_{\max})$. Cells above the diagonal are infeasible.
    This exploratory surface is not used for test-set selection; the final
    interval follows the dev-other protocol in
    \xref{Table}{tab:interval_selection}.}
    \label{fig:app_alpha_heatmaps}
\end{figure*}

\FloatBarrier
\section{Offline cost and measured W4 deployment}
\label{app:efficiency}

\xref{Figure}{fig:deployment_efficiency} summarizes our W4 deployment
measurements; \xrefrange{Tables}{tab:time}{tab:storage_memory} give the
per-model values and FADE's offline quantization cost.
Latency follows the ggml-CUDA/Q4\_0 protocol of TARQ~\citep{wang2026tarq} and
uses the ordinary W4 kernel. Each value is the median of three timed runs
after one warmup on an 11-second utterance. GPU runs use an A100-80GB; CPU
runs use eight threads on an AMD EPYC 7V12.

\begingroup
\setlength{\topsep}{2pt}

\begin{center}
\begin{minipage}{\columnwidth}
\centering
\captionof{table}{Offline quantization time (seconds; lower is better).}
\label{tab:time}
\normalsize
\setlength{\tabcolsep}{3.5pt}
\renewcommand{\arraystretch}{1.00}
\begin{tabular}{lrrr}
\toprule
Model & GPTQ & QEP & \methodname \\
\midrule
Whisper-Tiny & 19.09 & 25.82 & 30.60 \\
Whisper-Base & 32.33 & 43.10 & 52.03 \\
Whisper-Small & 80.21 & 103.94 & 127.01 \\
Qwen3-ASR-0.6B & 304.21 & 380.26 & 456.31 \\
Qwen3-ASR-1.7B & 924.79 & 1163.59 & 1404.30 \\
\bottomrule
\end{tabular}
\end{minipage}
\end{center}

\begin{center}
\begin{minipage}{\columnwidth}
\centering
\captionof{table}{Measured inference latency (seconds) for the same exported
models on an A100-80GB GPU and an eight-thread EPYC 7V12 CPU.}
\label{tab:inference_latency}
\footnotesize
\setlength{\tabcolsep}{4.2pt}
\renewcommand{\arraystretch}{1.02}
\begin{tabular}{lrrr}
\toprule
\multicolumn{4}{c}{A100-80GB} \\
\cmidrule(lr){1-4}
Model & FP16 & W4 & Speedup \\
\midrule
Whisper-Tiny & 0.89 & 0.84 & 1.06$\times$ \\
Whisper-Base & 0.92 & 0.84 & 1.10$\times$ \\
Whisper-Small & 1.06 & 0.91 & 1.16$\times$ \\
Qwen3-ASR-0.6B & 4.14 & 3.91 & 1.06$\times$ \\
Qwen3-ASR-1.7B & 8.99 & 4.12 & 2.18$\times$ \\
\bottomrule
\end{tabular}
\par\smallskip
\begin{tabular}{lrrr}
\toprule
\multicolumn{4}{c}{EPYC 7V12, eight threads} \\
\cmidrule(lr){1-4}
Model & FP16 & W4 & Speedup \\
\midrule
Whisper-Tiny & 0.52 & 0.34 & 1.53$\times$ \\
Whisper-Base & 1.05 & 0.62 & 1.69$\times$ \\
Whisper-Small & 3.68 & 1.78 & 2.07$\times$ \\
Qwen3-ASR-0.6B & 7.07 & 5.71 & 1.24$\times$ \\
Qwen3-ASR-1.7B & 13.36 & 9.32 & 1.43$\times$ \\
\bottomrule
\end{tabular}
\end{minipage}
\end{center}

Serialized W4 weights are 71.9\% smaller than FP16. Peak process memory
changes less because activations, runtime workspaces, and other model state
remain present.

\begin{center}
\begin{minipage}{\columnwidth}
\centering
\captionof{table}{Serialized weight storage and peak process memory (MB).
The final column in each group is the percentage saving or change from FP16.}
\label{tab:storage_memory}
\scriptsize
\setlength{\tabcolsep}{1.5pt}
\renewcommand{\arraystretch}{1.00}
\begin{tabular}{lrrrrrr}
\toprule
& \multicolumn{3}{c}{Weight storage} & \multicolumn{3}{c}{Peak process memory} \\
\cmidrule(lr){2-4}\cmidrule(lr){5-7}
Model & FP16 & W4 & $\Delta$\% & FP16 & W4 & $\Delta$\% \\
\midrule
Whisper-Tiny & 31.50 & 8.86 & $-$71.9 & 314.19 & 296.78 & $-$5.5 \\
Whisper-Base & 84.00 & 23.63 & $-$71.9 & 528.82 & 493.04 & $-$6.8 \\
Whisper-Small & 378.00 & 106.31 & $-$71.9 & 1467.60 & 1404.69 & $-$4.3 \\
Whisper-Large & 2800.00 & 787.50 & $-$71.9 & 8134.68 & 7706.70 & $-$5.3 \\
Qwen3-ASR-0.6B & 1170.80 & 329.27 & $-$71.9 & 3834.90 & 3835.95 & +0.0 \\
Qwen3-ASR-1.7B & 3264.00 & 918.00 & $-$71.9 & 9513.01 & 9550.14 & +0.4 \\
\bottomrule
\end{tabular}
\end{minipage}
\end{center}

\endgroup
\FloatBarrier

\end{document}